\documentclass[a4paper,11pt]{article}
\usepackage{a4wide}
\usepackage{amsfonts}
\usepackage{amsmath,amssymb}
\usepackage[mathscr]{eucal}
\newtheorem{theorem}{Theorem}
\newtheorem{proposition}{Proposition}
\newtheorem{definition}{Definition}
\begin{document}
\title{A modified Lax-Phillips scattering theory for quantum mechanics}
\author{Y. Strauss${}^{1}$ \\ Department of Mathematics, Ben-Gurion University of the Negev\\ Be'er Sheva 84105, Israel}
\maketitle
%
\setcounter{footnote}{1}
\footnotetext{e-mail: yossef.strauss@gmail.com}
\maketitle
\abstract{The Lax-Phillips scattering theory is an appealing abstract framework for the analysis of scattering resonances. Quantum mechanical adaptations of the theory have been proposed. However, since these quantum adaptations essentially retain the original structure of the theory, assuming the existence of incoming and outgoing subspaces for the evolution and requiring the spectrum of the generator of evolution to be unbounded from below, their range of applications is rather limited. In this paper it is shown that if we replace the assumption regarding the existence of incoming and outgoing subspaces by the assumption of the existence of Lyapunov operators for the quantum evolution (the existence of which has been proved for certain classes of quantum mechanical scattering problems) then it is possible to construct a structure analogous to the Lax-Phillips structure for scattering problems for which the spectrum of the generator of evolution is bounded from below.
\section{Introduction}
\par The Lax-Phillips scattering theory \cite{LP} had been originally developed for the analysis of scattering problems involving the solution of hyperbolic wave equations in domains exterior to  
compactly supported obstacles. As a theory formulated for such purposes, the Lax-Phillips theory, in its original form, is most suitable for dealing with resonances in the scattering of electromagnetic or 
acoustic waves off compact obstacles. The theory is based on a Hilbert space description of the propagating waves and the time evolution of these waves is given by a unitary evolution group.
\par Several aspects of the Lax-Phillips scattering theory distinguish it as an appealing abstract formalism for implementation even in situations outside of the strict range of problems for which it has 
been originally devised. The description of resonances in the framework of the Lax-Phillips theory possesses properties which may be considered as defining properties of an appropriate description of these 
objects. One such property is a dynamical characterization of resonances via their time evolution given in terms of a continuous, one parameter, strongly contractive semigroup known as the Lax-Phillips 
semigroup. Specifically, resonances are identified as eigenvalues of the generator of the Lax-Phillips semigroup. This corresponds to another desirable feature of the theory, namely, the fact that each 
resonance pole is associated with a resonance state (or more generally a subspace) in a Hilbert space. In fact, the Lax-Phillips semigroup is obtained by a projection of the unitary evolution of the full system onto the subspace spanned by 
resonance states.
\par Consider a Hilbert space $\mathcal H^{\text{\tiny LP}}$ and a continuous, one parameter, evolution group of unitary operators $\{U(t)\}_{t\in\mathbb R}$ on $\mathcal H^{\text{\tiny LP}}$. The starting 
point for the Lax-Phillips scattering theory is the assumption that there exist in $\mathcal H^{\text{\tiny LP}}$ two distinguished subspaces $\mathcal D_-$ and $\mathcal D_+$ with the properties
\begin{eqnarray}
\label{LP_assumptions}
 \mathcal D_- & \perp &\mathcal D_+ \nonumber\\
 U(t)\mathcal D_- &\subseteq & \mathcal D_-,\quad \forall t\leq 0 \nonumber\\
 U(t)\mathcal D_+ &\subseteq & \mathcal D_+,\quad \forall t\geq 0\\
 \cap_{t\in\mathbb R}U(t)\mathcal D_\pm &=& \{0\} \nonumber\\
 \overline{\vee_{t\in\mathbb R}U(t)\mathcal D_\pm} &=& \mathcal H^{\text{\tiny LP}} \nonumber
\end{eqnarray}
We call a Hilbert space $\mathcal H^{\text{\tiny LP}}$ on which the assumptions in Eq. (\ref{LP_assumptions}) hold a Lax-Phillips Hilbert space. The subspaces $\mathcal D_-$ and $\mathcal D_+$ are called, respectively, the \emph{incoming subspace} and \emph{outgoing subspace} for the evolution $\{U(t)\}_{t\in\mathbb R}$. The subspace $\mathcal D_-$ 
corresponds to incoming waves which do not interact with the target prior to $t=0$ and the subspace $\mathcal D_+$ corresponds to outgoing waves which do not interact with the target after $t=0$. These 
properties are reflected in the stability properties of $\mathcal D_-$ and $\mathcal D_+$ in Eq. (\ref{LP_assumptions}) above.
\par Let $P_-$ be the orthogonal projection in $\mathcal H^{\text{\tiny LP}}$ onto the orthogonal complement of $\mathcal D_-$ and $P_+$ be the orthogonal projection in $\mathcal H^{\text{\tiny LP}}$ onto the orthogonal complement 
of $\mathcal D_+$. The main object of study in the Lax-Phillips theory is the family $\{Z_{\text{\tiny LP}}(t)\}_{t\geq 0}$ of operators on $\mathcal H^{\text{\tiny LP}}$ defined by
\begin{equation}
\label{LP_semigroup_def}
 Z_{\text{\tiny LP}}(t):=P_+U(t)P_-,\quad t\geq 0.
\end{equation}
Lax and Phillips prove the following theorem:
\begin{theorem}
The operators $Z_{\text{\tiny LP}}(t)$, $t\geq 0$, annihilate $\mathcal D_+$ and $\mathcal D_-$, map the orthogonal complement subspace $\mathcal H^{\text{\tiny LP}}_{res}:=\mathcal H^{\text{\tiny LP}}\ominus(\mathcal D_-\oplus\mathcal D_+)$ into itself and form a strongly continuous semigroup (i.e., $Z_{\text{\tiny LP}}(t_1)Z_{\text{\tiny LP}}(t_2)=Z_{\text{\tiny LP}}(t_1+t_2)$, $t_1,t_2\geq 0$) of contraction operators on $\mathcal H^{\text{\tiny LP}}_{res}$. Furthermore, we have $s-\lim_{t\to\infty} Z_{\text{\tiny LP}}(t)=0$.\hfill$\square$
\end{theorem}
The family of operators $\{Z_{\text{\tiny LP}}(t)\}_{t\geq 0}$ is known as the \emph{Lax-Phillips semigroup}.
\par Let $L^2(\mathbb R,\mathcal K)$ denote the space of Lebesgue square integrable functions defined on the real line $\mathbb R$ and taking their values in a separable Hilbert space $\mathcal K$. Ja. G. 
Sinai \cite{CFS} proved that if the assumptions in Eq. (\ref{LP_assumptions}) hold for the outgoing subspace $\mathcal D_+$ then the following theorem holds:
\begin{theorem}[(Ja. G. Sinai)]
If $\mathcal D_+$ is an outgoing subspace with respect to the unitary group $\{U(t)\}_{t\in\mathbb R}$ defined on a Hilbert space $\mathcal H^{\text{\tiny LP}}$ then $\mathcal H^{\text{\tiny LP}}$ can be represented isometrically as the Hilbert space of functions $L^2(\mathbb R,\mathcal K)$ for some Hilbert space $\mathcal K$ (called the auxiliary Hilbert space) in such a way that $U(t)$ goes to translation to the right by $t$ units and $\mathcal D_+$ is mapped onto $L^2(\mathbb R_+,\mathcal K)$. This representation is unique up to an isomorphism of $\mathcal K$.\hfill$\square$
\end{theorem}
A representation of this kind is called an \emph{outgoing translation representation}. An analogous representation theorem holds for an incoming subspace $\mathcal D_-$, i.e., if $\mathcal D_-$ is an 
incoming subspace with respect to the group $\{U(t)\}_{t\in\mathbb R}$ then there is a representation in which $\mathcal H^{\text{\tiny LP}}$ is mapped isometrically onto $L^2(\mathbb R,\mathcal K)$, $U(t)$
goes to translation to the right by $t$ units and $\mathcal D_-$ is mapped onto $L^2(\mathbb R_-,\mathcal K)$. This representation is called an \emph{incoming translation representation}. 
\par Let $W^{\text{\tiny LP}}_+\,:\mathcal H^{\text{\tiny LP}}\mapsto L^2(\mathbb R,\mathcal K)$ and $W^{\text{\tiny LP}}_-\,:\,\mathcal H^{\text{\tiny LP}}\mapsto L^2(\mathbb R,\mathcal K)$ be the mappings of $\mathcal H^{\text{\tiny LP}}$ onto the outgoing and incoming translation representations respectively. The map $S_{\text{\tiny LP}}\,:\,L^2(\mathbb R,\mathcal K)\mapsto L^2(\mathbb R,\mathcal K)$ defined by
\begin{equation*}
 S_{\text{\tiny LP}}:=W^{\text{\tiny LP}}_+\left(W^{\text{\tiny LP}}_-\right)^{-1}
\end{equation*}
is called the \emph{Lax-Phillips scattering operator}. It was proved by Lax and Phillips that $S_{\text{\tiny LP}}$ is equivalent to the standard definition of the scattering operator. For most purposes it is more convenient not to work with the incoming and outgoing translation representations but rather with their Fourier transforms called, respectively, the \emph{incoming spectral representation} and 
\emph{outgoing spectral representation}. According to the Paley-Wiener theorem \cite{PW} in the incoming spectral representation $\mathcal D_-$ is represented by $\mathcal H^2_+(\mathbb R,\mathcal K)$ where $\mathcal H^2_+(\mathbb R,\mathcal K)$ is the space of boundary values on $\mathbb R$ of functions in the Hardy space $\mathcal H^2(\mathbb C^+,\mathcal K)$ of vector valued functions (with values in 
$\mathcal K$) defined on the upper half-plane $\mathbb C^+$. By the same theorem in the outgoing spectral representation $\mathcal D_+$ is represented by $\mathcal H^2_-(\mathbb R,\mathcal K)$ where 
$\mathcal H^2_-(\mathbb R,\mathcal K)$ is the space of boundary values on $\mathbb R$ of functions in the Hardy space $\mathcal H^2(\mathbb C^-,\mathcal K)$ of vector valued functions (with values in 
$\mathcal K$) defined on the lower half-plane $\mathbb C^-$. The transformation to the spectral representations implies a transformation of the scattering operator $S_{\text{\tiny LP}}$ into the scattering operator in the spectral representation $\hat{\mathcal S}_{\text{\tiny LP}}$  defined by
\begin{equation*}
 \hat S_{\text{\tiny LP}}:=FS_{\text{\tiny LP}}F^{-1}
\end{equation*}
where $F$ is the Fourier transform operator. The operator $\hat S_{\text{\tiny LP}}$ is then realized in the spectral representation as a multiplicative, operator valued function 
$\hat{\mathcal S}_{\text{\tiny LP}}(\cdot)\,:\,\mathbb R\mapsto \mathscr B(\mathcal K)$ (where $\mathscr B(\mathcal K)$ is the space of all bounded operators on $\mathcal K$) having the properties:
\begin{description}
\item{(a)} $\hat S_{\text{\tiny LP}}(\cdot)$ is the boundary value on $\mathbb R$ of an operator valued function $\hat{\mathcal S}_{\text{\tiny LP}}(\cdot)\,:\,\mathbb C^+\mapsto \mathscr B(\mathcal K)$ analytic on $\mathbb C^+$,
\item{(b)} $\Vert\hat{\mathcal S}_{\text{\tiny LP}}(z)\Vert\leq 1$, $\forall z\in\mathbb C^+$,
\item{(c)} $\hat S_{\text{\tiny LP}}(E)$, $E\in\mathbb R$ is, pointwise, a unitary operator on $\mathcal K$.
\end{description}
The operator valued function $\hat{\mathcal S}_{\text{\tiny LP}}(\cdot)$ is called the \emph{Lax-Phillips S-matrix}, this function is characterized by its action on the subspace 
$\mathcal H^2_+(\mathbb R,\mathcal K)$ as being an \emph{inner (operator valued) function} \cite{SzNF,RR,Hof}. The analytic continuation of $\hat{\mathcal S}_{\text{\tiny LP}}(\cdot)$ from the upper half-plane 
to the lower half-plane is given by
\begin{equation*}
  \hat{\mathcal S}_{\text{\tiny LP}}(z):=[\hat{\mathcal S}_{\text{\tiny LP}}^*(\overline z)]^{-1},\quad \text{Im }z<0
\end{equation*}
It can then be shown that the analytic continuation of the Lax-Phillips $S$-matrix to the whole complex plane is a meromorphic operator valued function. One of the main results of the Lax-Phillips scattering theory is:
\begin{theorem}
\label{LP_main_result}
Let $B$ denote the generator of the semigroup $\{Z_{\text{\tiny LP}}(t)\}_{t\geq 0}$. If $\text{Im }\mu<0$, then $\mu$ belongs to the point spectrum of $B$ if and only if $\hat{\mathcal S}_{\text{\tiny LP}}^*(\overline\mu)$ has a non-trivial null space.\hfill$\square$
\end{theorem}
Theorem \ref{LP_main_result} implies that a pole of the Lax-Phillips $S$-matrix at a point $\mu$ in the lower half plane is associated with an eigenvalue $\mu$ of the generator of the Lax-Phillips semigroup.
In other words, resonance poles of the Lax-Phillips $S$-matrix correspond to eigenvalues of the (generator of the) Lax-Phillips semigroup with well defined eigenvectors belonging to the resonance subspace 
$\mathcal H^{\text{\tiny LP}}_{res}=\mathcal H^{\text{\tiny LP}}\ominus(\mathcal D_+\oplus\mathcal D_-)$. 
\par The attractive properties of the Lax-Phillips scattering theory, mentioned above, have led to some efforts to adapt the formalism into the framework of quantum mechanics. Early work in this 
direction can be found, for example, in Refs. \cite{Pav1,Pav2,FP,HP,EH} (see also Ref. \cite{KMPY} for a more recent application of the Lax-Phillips structure to quantum problems). A general formalism was 
developed in Ref. \cite{SHE} and subsequently applied to several physical models in Refs. \cite{SH1,SH2,BAriH}. However, in general 
one cannot apply, without modification, the basic structure of the Lax-Phillips scattering theory in the context of standard quantum mechanical scattering problems since incoming and outgoing subspaces $\mathcal D_\pm$ having the 
properties listed in Eq. (\ref{LP_assumptions}) cannot be found for large classes of such problems. This can be seen, for example, by noting the fact that in the Lax-Phillips theory the continuous 
spectrum of the generator of evolution is necessarily unbounded from below as well as from above, a requirement which is not met by most quantum mechanical Hamiltonians. Hence, the range of applications of quantum mechanical adaptations of the 
Lax-Phillips theory which essentially retain the original mathematical structure of the theory is rather limited.
\par A step forward in the efforts to approximate the structure of the Lax-Phillips theory within the context of quantum mechanics has been made in Ref. \cite{S1} with the introduction into the 
framework of quantum mechanics of \emph{forward and backward Lyapunov operators}, based on properties of Hardy spaces, and subsequent investigation of their properties and 
their applications in Refs. \cite{S2,SSMH1,SSMH2}. If $\mathcal H$ is the Hilbert space corresponding to a given system and $H$ is a self-adjoint generator of evolution of the system we define the trajectory $\Phi_\varphi$ corresponding to a state $\varphi\in\mathcal H$ to be the set of states
\begin{equation*}
 \Phi_\varphi:=\{\varphi(t)\}_{t\in\mathbb R}=\{U(t)\varphi\}_{t\in\mathbb R},
\end{equation*}
where $U(t)=\exp(-iHt)$. Note that this definition extends the definition of trajectory in Ref. \cite{S1} to include negative as well as positive times. Accordingly, the definition of a 
\emph{forward Lyapunov operator} in Ref. \cite{S1} is also extended as follows:
\begin{definition}[forward Lyapunov operator]
Let $M$ be a bounded self-adjoint operator on $\mathcal H$. Let $\Phi_\varphi$ be the trajectory corresponding to an arbitrarily chosen normalized 
state $\varphi\in\mathcal H$. Let  
$M(\Phi_\varphi):=\{(\psi,M\psi)\mid \psi\in\Phi_\varphi\}$ be the collection of all expectation values of $M$ for states in $\Phi_\varphi$. Then $M$ is a forward Lyapunov operator if the mapping $\tau_{M,\varphi}\,:\,\mathbb R\mapsto M(\Phi_\varphi)$ defined by
\begin{equation*}
 \tau_{M,\varphi}(t)=(\varphi(t),M\varphi(t))
\end{equation*}
is one to one and monotonically decreasing.\hfill$\square$
\end{definition}
\begin{description}
\item{\bf Remark 1:} We assume throughout the present paper that all generators of evolution are time independent and, therefore, we have symmetry of the evolution with respect to time translations.
\item{\bf Remark 2:} If in the definition above we require that $\tau_{M,\varphi}$ be monotonically increasing instead of monotonically decreasing we also obtain a valid definition of a forward Lyapunov 
operator. The requirement that $\tau_{M,\varphi}$ is monotonically decreasing is made purely for the sake of convenience.
\end{description}
If $M$ is a forward Lyapunov operator then we are able to find the time ordering of states in the trajectory $\Phi_\varphi$ according to the ordering of expectation values in $M(\Phi_\varphi)$. Hence, the 
existence of a Lyapunov operator introduces temporal ordering into the Hilbert space $\mathcal H$ of a problem for which such an operator can be constructed. The definition of a backward Lyapunov operator 
is similar to that of a forward Lyapunov operator, but with respect to the reversed direction of time. The significance of the existence of forward and backward Lyapunov operators in the construction of a 
formalism analogous to the Lax-Phillips theory within quantum mechanics can be understood if we consider again the original Lax-Phillips formalism, and, in particular, the properties of the projection 
operators $P_+$ and $P_-$. In fact, from the representation of $P_+$ in the outgoing translation representation as an orthogonal projection on the subspace $L^2(\mathbb R_-;\mathcal K)$ (or, indeed, 
directly from the  definition of $P_+$ and the properties of $\mathcal D_\pm$ in Eq. (\ref{LP_assumptions})) it is evident that $P_+$ is a forward Lyapunov operator for the evolution in the Lax-Phillips 
theory. For every $\psi\in \mathcal H^{\text{\tiny LP}}$ we have
\begin{equation*}
  (\psi(t_2),P_+\psi(t_2))\leq(\psi(t_1),P_+\psi(t_1)),\quad t_1\leq t_2,\qquad \lim_{t\to\infty}(\psi(t),P_+\psi(t))=0. 
\end{equation*}
Likewise, from the representation of $P_-$ in the incoming translation representation as an orthogonal projection on the subspace $L^2(\mathbb R_+;\mathcal K)$ it is evident that $P_-$ is a backward Lyapunov operator for the Lax-Phillips evolution satisfying
\begin{equation*}
  (\psi(t_2),P_-\psi(t_2))\leq(\psi(t_1),P_-\psi(t_1)),\ t_2\leq t_1,\qquad \lim_{t\to -\infty}(\psi(t),P_-\psi(t))=0.
\end{equation*}
Note that, if we define $Z(t):=P_+U(t)$ for $t\geq 0$ then, by the stability properties of $\mathcal D_+$, we have for $t_1,t_2\geq 0$
\begin{equation*}
 Z(t_1)Z(t_2)=P_+U(t_1)P_+U(t_2)=P_+U(t_1)(P_+ +P_+^\perp)U(t_2)=P_+U(t_1+t_2)=Z(t_1+t_2),
\end{equation*}
where $P_+^\perp=I-P_+$. Hence, the family of operators $\{Z(t)\}_{t\geq 0}$ is a continuous, one parameter, contractive semigroup on $\mathcal H^{\text{\tiny LP}}$. It is easy to show, in addition, that 
$s-\lim_{t\to\infty}Z(t)=0$. Moreover, we have
\begin{equation}
\label{LP_intertwining_rel}
 P_+U(t)=P_+U(t)(P_+ +P_+^\perp)=P_+U(t)P_+=Z(t)P_+,\quad t\geq 0,
\end{equation}
so that for non-negative times $P_+$ intertwines the unitary evolution $U(t)$ with the semigroup evolution $Z(t)$. Finally, observe that by the intertwining relation in Eq. (\ref{LP_intertwining_rel}) we have
\begin{equation}
 \label{LP_semigroup_alternate_rep}
 Z_{\text{\tiny LP}}(t)=P_+U(t)P_- =Z(t)P_+P_-,\quad t\geq 0
\end{equation}
\par We turn now to consider Lyapunov operators in quantum mechanics. Following the basic existence results proved in Ref. \cite{S1} it has been shown in Refs. \cite{S2,SSMH1,SSMH2} that if a quantum 
mechanical scattering problem satisfies the assumptions that: (a) The absolutely continuous spectrum of the unperturbed and perturbed Hamiltonians is $\sigma_{ac}(H_0)=\sigma_{ac}(H)=\mathbb R^+$,(b) The 
multiplicity of the a.c. spectrum of $H$ is uniform, (c) The incoming and outgoing M\o ller wave operators $\Omega_\pm(H_0,H)$ exist and are complete; then, if $\mathcal H_{ac}$ is the subspace of 
$\mathcal H$ corresponding to the a.c. spectrum of $H$, there exists a self-adjoint, contractive, injective and non-negative forward Lyapunov operator 
$M_+\,:\,\mathcal H_{ac}\mapsto\mathcal H_{ac}$ for the quantum evolution, i.e., for any $\psi\in\mathcal H_{ac}$ we have
\begin{equation*}
 (\psi(t_2),M_+\psi(t_2))\leq (\psi(t_1),M_+\psi(t_1)),\quad t_1\leq t_2,\qquad \lim_{t\to\infty}(\psi(t),M_+\psi(t))=0
\end{equation*}
where $\psi(t)=U(t)\psi=\exp(-iHt)\psi$. In addition, it is shown in Refs. \cite{S2,SSMH1,SSMH2} that  $\text{Ran }M_+$ is dense in $\mathcal H_{ac}$. Similarly, under the same assumptions, there exists a 
self-adjoint, contractive, injective and non-negative backward Lyapunov operator $M_-\,:\,\mathcal H_{ac}\mapsto\mathcal H_{ac}$ for the quantum evolution
\begin{equation*}
 (\psi(t_2),M_-\psi(t_2))\leq (\psi(t_1),M_-\psi(t_1)),\quad t_2\leq t_1,\qquad \lim_{t\to-\infty}(\psi(t),M_-\psi(t))=0,
\end{equation*}
with $\text{Ran }M_-$ dense in $\mathcal H_{ac}$.
\par Set $\Lambda_+:=M_+^{1/2}$ and $\Lambda_-:=M_-^{1/2}$. Upon comparison to the definition of the 
Lax-Phillips semigroup in Eq. (\ref{LP_semigroup_def}) we are led to define for a quantum mechanical scattering problem a family of operators 
$\{Z_{app}(t)\}_{t\geq 0}\,:\,\mathcal H_{ac}\mapsto\mathcal H_{ac}$, to which we refer as the \emph{approximate Lax-Phillips semigroup}, via the definition
\begin{equation}
\label{approx_lax_phillips_semigroup}
 Z_{app}(t):=\Lambda_+U(t)\Lambda_-,\quad t\geq 0.
\end{equation}
Note that if we apply similar definitions of $\Lambda_\pm$, as square roots of the Lyapunov operators, in the Lax-Phillips case we obtain $\Lambda_+=P_+^{1/2}=P_+$ and $\Lambda_- = P_-^{1/2}=P_-$ so that
in this case we have $Z_{app}(t)=Z_{\text{\tiny LP}}(t)$, $\forall t\geq 0$.
\par It is shown in Refs. \cite{S2,SSMH1} that there exists a continuous, strongly contractive, one parameter semigroup $\{Z_+(t)\}_{t\geq 0}$ such that for each $\psi\in\mathcal H_{ac}$ we have
\begin{equation*}
 \Vert Z_+(t_2)\psi\Vert\leq\Vert Z_+(t_1)\psi\Vert,\quad t_2\geq t_1\geq 0,\qquad s-\lim_{t\to\infty}Z_+(t)=0,
\end{equation*}
and the following intertwining relation holds
\begin{equation}
\label{QM_forward_intertwining_rel}
  \Lambda_+U(t)=Z_+(t)\Lambda_+,\quad U(t)=e^{-iHt},\ t\geq 0
\end{equation}
(a similar semigroup and intertwining relation can be found for $\Lambda_-$ in the backward direction of time). Using this intertwining relation we obtain
\begin{equation}
\label{approx_semigroup_alternate_rep}
 Z_{app}(t):=\Lambda_+U(t)\Lambda_- =Z_+(t)\Lambda_+\Lambda_-,\quad t\geq 0.
\end{equation}
Eq. (\ref{QM_forward_intertwining_rel}) is to be compared with Eq. (\ref{LP_intertwining_rel}) and Eq. (\ref{approx_semigroup_alternate_rep}) is to be compared with Eq. (\ref{LP_semigroup_alternate_rep}). Note, however, that $\{Z_{app}(t)\}_{t\in\mathbb R^+}$ is not an exact semigroup.
\par For a quantum mechanical scattering problem satisfying assumptions (a)-(c) the scattering operator $S_{\text{\tiny QM}}=\Omega_-^{-1}\Omega_+$, where $\Omega_+$ and $\Omega_-$ are, respectively, the incoming and outgoing M\o ller wave operators, has a representation as a mapping from the incoming energy representation to the outgoing energy representation in terms
of the scattering matrix $\hat S_{\text{\tiny QM}}(\cdot): \mathbb R^+\mapsto\mathscr U(\mathcal K)$, where $\mathscr U(\mathcal K)$ is the set of unitary 
operators on the multiplicity Hilbert space $\mathcal K$ (note that $\hat S_{\text{\tiny QM}}(\cdot)$ is, in fact, a representation of the scattering operator 
$S_{\text{\tiny QM}}$ in the spectral representation of the unperturbed Hamiltonian $H_0$). The scattering matrix $\hat S_{\text{\tiny QM}}(\cdot)$ in the quantum case is analogous 
to the Lax-Phillips scattering martrix $\hat S_{\text{\tiny LP}}(\cdot)$ in the Lax-Phillips case, which is also a mapping between incoming and outgoing spectral 
representations of the generator of evolution. Adding the correspondence of these two objects to the list of analogous constructions for the 
quantum machanical scattering theory and the Lax-Phillips scattering theory discussed above we may produce, for a quantum mechanical scattering 
problem satisfying assumptions (a)-(c), the following list of correspondences between objects in the Lax-Phillips scattering theory and corresponding objects in the case of quantum mechanical scattering
\begin{eqnarray}
\label{lp_qm_analogy_ver_1}
 \underline{\text{LP scattering theory}} &{ }&  \underline{\text{QM scattering theory}}\notag\\
 U(t)=e^{-iKt} &\Longleftrightarrow& U(t)=e^{-iHt}\notag\\
 P_\pm &\Longleftrightarrow& \Lambda_\pm\\
 Z_{\text{\tiny{LP}}}(t)=P_+U(t)P_-,\ t\geq 0 &\Longleftrightarrow& Z_{app}(t)=\Lambda_+U(t)\Lambda_-,\ t\geq 0\notag\\
 \hat S_{\text{\tiny LP}}(E),\ \ E\in\mathbb R &\Longleftrightarrow& \hat S_{\text{\tiny QM}}(E),\ \  E\in\mathbb R^+\notag
\end{eqnarray}
\par Our goal in the present paper is to construct, in the context of quantum mechanical scattering, a formalism analogous to the Lax-Phillips
scattering theory. Thus far we have considered in the quantum mechanical case a set of objects analogous to the central objects 
of the Lax-Phillips theory. However, beyond analogy in the construction of certain objects what we seek for is a result analogous to Theorem \ref{LP_main_result}, the central 
theorem of the Lax-Phillips scattering theory associating resonance poles of the Lax-Phillips S-matrix to eigenvalues and eigenfunctions of the 
Lax-Phillips semigroup. Note that we cannot expect to obtain in the quantum mechanical case an exact parallel of Theorem \ref{LP_main_result} since, as mentioned above, $\{Z_{app}(t)\}_{t\in\mathbb R^+}$ 
is not an exact semigroup. The task of proving a theorem analogous to Theorem \ref{LP_main_result} in the case of quantum mechanical scattering processes, at least in an appropriately defined approximate 
sense, is taken up in Section \ref{res_poles_and_states_and_approx_LP_semigroup} and Section \ref{proof_of_main_theorem} where it is proved that to a resonance pole of the quantum mechanical 
S-matrix $\hat S_{\text{\tiny QM}}(\cdot)$ in the second sheet of the complex energy Riemann surface, at a point $\mu$ with $\text{Im }\mu<0$, there corresponds a state $\psi_\mu^{res}$ which is an 
approximate eigenfunction of each element of $\{Z_{app}(t)\}_{t\in\mathbb R^+}$ in the sense that
\begin{equation*}
 Z_{app}(t)\psi_\mu^{res}=e^{-i\mu t}\psi_\mu^{res}+\text{small corrections},\quad t\geq 0.
\end{equation*}
The state $\psi_\mu^{res}$ is considered to be a resonance state corresponding to the resonance pole of $\hat S_{\text{\tiny QM}}(\cdot)$ at $z=\mu$. By establishing a result analogous to Theorem \ref
{LP_main_result} in the context of quantum mechanical resonance scattering (in an appropriately defined approximate sense) we complete the construction of a framework analogous to the Lax-Phillip theory for 
quantum mechanical scattering. We note that $\psi_\mu^{res}$ is an exact eigenstate of each element of the semigroup $\{Z_+(t)\}_{t\geq 0}$ in the same way that 
a resonance state in the Lax-Phillips theory is an eigenstate of each element of the semigroup $\{Z(t)\}_{t\geq 0}$ .
\par The arrangement of the rest of the present paper is as follows: Section \ref{Lyapunov_op_trans_rep_in_LP_QM} provides a detailed discussion of Lyapunov operators for the case of a scattering problem 
satisfying assumptions (i)-(ii) below. As mentioned above, since incoming and outgoing subspaces $\mathcal D_\pm$ cannot be found, in general, for standard quantum mechanical scattering problems, and hence
the construction of a formalism analogous to the Lax-Phillips scattering theory cannot be based on the existence of such subspaces, the basic objects involved in the construction of a structure 
approximating the Lax-Phillips structure in the quantum case are the Lyapunov operators analogous to the Lyapunov operators $P_\pm$ of the Lax-Phillips theory. The existence of the Lyapunov 
operators in the quantum case leads to the definition of objects (such as the approximate Lax-Phillips semigroup) and representations (called the forward and backward transition representations) analogous 
to the objects and representations of the Lax-Phillips theory. These are also discussed in Section \ref{Lyapunov_op_trans_rep_in_LP_QM}. Section \ref{res_poles_and_states_and_approx_LP_semigroup} is 
centered on the discussion, in the quantum mechanical context, of a result analogous to Theorem \ref{LP_main_result}, the main result of the Lax-Phillips theory associating with each resonance pole of the 
Lax-Phillips S-matrix a resonance state in the Lax-Phillips Hilbert space $\mathcal H^{\text{\tiny LP}}$. An analogous (approximate) result in the quantum mechanical case is given by 
Theorem \ref{approx_proj_estimate} in Section \ref{res_poles_and_states_and_approx_LP_semigroup}. The proof of Theorem \ref{approx_proj_estimate} is contained in Section \ref{proof_of_main_theorem}. 
Conclusions are given in Section \ref{conclusions}.
\section{Lyapunov operators and transition representations in Lax-Phillips theory and in quantum mechanics}
\label{Lyapunov_op_trans_rep_in_LP_QM}
\par Let $\mathcal K$ be a separable Hilbert space and let $L^2(\mathbb R;\mathcal K)$ be the Hilbert space of Lebesgue square integrable $\mathcal K$ valued functions defined on $\mathbb R$. 
Let $\hat E$ be the operator of multiplication by the independent variable on $L^2(\mathbb R;\mathcal K)$. Let $\{u(t)\}_{t\in\mathbb R}$ be the continuous, one parameter, unitary evolution group on $L^2(\mathbb R;\mathcal K)$ generated by $\hat E$, i.e.,
\begin{equation}
\label{u_evolution_group}
 [u(t)f](E)=[e^{-i\hat Et}f](E)=e^{-iEt}f(E),\quad f\in L^2(\mathbb R;\mathcal K),\ \ E\in\mathbb R.
\end{equation}
Let $\mathcal H^2(\mathbb C^+;\mathcal K)$ and $\mathcal H^2(\mathbb C^-;\mathcal K)$ be, respectively, the Hardy space of $\mathcal K$ valued functions analytic in $\mathbb C^+$ and $\mathbb C^-$. As mentioned in the introduction, the Hilbert space $\mathcal H^2_+(\mathbb R;\mathcal K)$ consisting of nontangential boundary values 
on the real axis of functions in $\mathcal H^2(\mathbb C^+;\mathcal K)$ is isomorphic to $\mathcal H^2(\mathbb C^+;\mathcal K)$. Similarly, the Hilbert space $\mathcal H^2_-(\mathbb R;\mathcal K)$ of 
non-tangential boundary value functions of functions in $\mathcal H^2(\mathbb C^-;\mathcal K)$ is isomorphic to $\mathcal H^2(\mathbb C^-;\mathcal K)$. The spaces $\mathcal H^2_\pm(\mathbb R;\mathcal K)$ 
are orthogonal subspaces of $L^2(\mathbb R;\mathcal K)$ and we have
\begin{equation*}
 L^2(\mathbb R;\mathcal K)=\mathcal H^2_+(\mathbb R;\mathcal K)\oplus \mathcal H^2_-(\mathbb R;\mathcal K).
\end{equation*}
We denote the orthogonal projections in $L^2(\mathbb R;\mathcal K)$ on $\mathcal H^2_+(\mathbb R;\mathcal K)$ and $\mathcal H^2_-(\mathbb R;\mathcal K)$, respectively, by $\hat P_+$ and $\hat P_-$.
\par Recall that in the Lax-Phillips theory $P_+$ is the orthogonal projection on the orthogonal complement of the outgoing subspace $\mathcal D_+$. In the outgoing translation representation the 
Lax-Phillips Hilbert space $\mathcal H^{\text{\tiny LP}}$ is mapped isometrically onto the function space $L^2(\mathbb R;\mathcal K)$ where $\mathcal K$ is the auxiliary Hilbert space and $\mathcal D_+$ is mapped onto 
the subspace $L^2(\mathbb R_+;\mathcal K)$. The evolution $U(t)$ is represented by translation to the right by $t$ units. The outgoing spectral representation is a spectral representation of the generator 
$K$ of the unitary evolution group $\{U(t)\}_{t\in\mathbb R}=\{\exp(-iKt)\}_{t\in\mathbb R}$ of the Lax-Phillips theory and is obtained by Fourier transform of the outgoing translation representation. In 
this representation $\mathcal H^{\text{\tiny LP}}$ is represented by $L^2(\mathbb R;\mathcal K)$, the evolution group $\{U(t)\}_{t\in\mathbb R}$ is represented by the group $\{u(t)\}_{t\in\mathbb R}$ in Eq. (\ref{u_evolution_group}) and, by the Paley-Wiener theorem, $\mathcal D_+$ is represented by the Hardy space $\mathcal H^2_-(\mathbb R;\mathcal K)$. Therefore $P_+$ is represented in this representation by the projection $\hat P_+$ on $\mathcal H^2_+(\mathbb R;\mathcal K)$. Hence if $\hat W^{\text{\tiny LP}}_+\,:\,\mathcal H^{\text{\tiny LP}}\mapsto L^2(\mathbb R;\mathcal K)$ is the mapping of $\mathcal H^{\text{\tiny LP}}$ onto the outgoing spectral representation we have
\begin{equation}
\label{LP_forward_lyapunov_op_construction}
 P_+ =\left(\hat W^{\text{\tiny LP}}_+\right)^{-1}\hat P_+ \hat W^{\text{\tiny LP}}_+.
\end{equation}
\par In a similar manner, in the incoming translation representation the Lax-Phillips Hilbert space $\mathcal H^{\text{\tiny LP}}$ is mapped isometrically onto the function space $L^2(\mathbb R;\mathcal K)$
and the incoming subspace $\mathcal D_-$ is mapped onto the subspace $L^2(\mathbb R_-;\mathcal K)$. The evolution $U(t)$ is again represented by translation to the right by $t$ units. The incoming spectral 
representation, obtained by Fourier transform of the incoming translation representation, is a spectral representation of the generator $K$ of the evolution group $\{U(t)\}_{t\in\mathbb R}$. In this 
representation $\mathcal H^{\text{\tiny LP}}$ is represented by the function space $L^2(\mathbb R;\mathcal K)$, the evolution group $\{U(t)\}_{t\in\mathbb R}$ is represented by the group $\{u(t)\}_{t\in\mathbb R}$ of Eq. (\ref{u_evolution_group}) and, 
by the Paley-Wiener theorem, $\mathcal D_-$ is represented by the Hardy space $\mathcal H^2_+(\mathbb R;\mathcal K)$. Therefore $P_-$, the projection on the orthogonal complement of $\mathcal D_-$, is 
represented in the incoming spectral representation by the projection $\hat P_-$ on $\mathcal H^2_-(\mathbb R;\mathcal K)$ and hence, if $\hat W^{\text{\tiny LP}}_-\,:\,\mathcal H^{\text{\tiny LP}}\mapsto L^2(\mathbb R;\mathcal K)$ is the mapping of $\mathcal H^{\text{\tiny LP}}$ onto the incoming spectral representation, we have
\begin{equation}
\label{LP_backward_lyapunov_op_construction}
 P_- =\left(\hat W^{\text{\tiny LP}}_-\right)^{-1}\hat P_- \hat W^{\text{\tiny LP}}_-.
\end{equation}
Observe that Eqns. (\ref{LP_forward_lyapunov_op_construction}) and (\ref{LP_backward_lyapunov_op_construction}) provides us with an explicit procedure for the construction of the forward and backward 
Lyapunov operators $P_\pm$ in the Lax-Phillips theory.
\par Next, we turn to consider the construction of Lyapunov operators in the quantum mechanical case. Let $L^2(\mathbb R_\pm;\mathcal K)$ be the subspaces of $L^2(\mathbb R;\mathcal K)$ consisting of functions supported on $\mathbb R_\pm$. Then we have another orthogonal decomposition of $L^2(\mathbb R;\mathcal K)$
\begin{equation*}
 L^2(\mathbb R;\mathcal K)=L^2(\mathbb R_+;\mathcal K)\oplus L^2(\mathbb R_-;\mathcal K).
\end{equation*}
We denote the orthogonal projections on the subspaces $L^2(\mathbb R_+;\mathcal K)$ and $L^2(\mathbb R_-;\mathcal K)$, respectively, by $P_{\mathbb R_+}$ and $P_{\mathbb R_-}$. Let $\hat E_+$ be the 
operator of multiplication by the independent variable on $L^2(\mathbb R_+;\mathcal K)$. Let $\{u_+(t)\}_{t\in\mathbb R}$ be the continuous, one parameter, unitary evolution group generated by $\hat E_+$, 
i.e.,
\begin{equation*}
 [u_+(t)f](E)=[e^{-i\hat E_+ t}f](E)=e^{-iEt}f(E),\quad f\in L^2(\mathbb R^+;\mathcal K),\quad E\in\mathbb R_+.
\end{equation*}
The following two theorems, first proved in Ref. \cite{S1}, form the basis for the present discussion of Lyapunov operators and their applications in the context of quantum mechanical scattering: 
\begin{theorem}
\label{mf_theorem}
Let $M_F: L^2(\mathbb R^+;\mathcal K)\mapsto L^2(\mathbb R^+;\mathcal K)$ be the operator defined by
\begin{equation}
\label{mf_definition}
 M_F:=(P_{\mathbb R^+}\hat P_+P_{\mathbb R^+})\vert_{L^2(\mathbb R^+;\mathcal K)}.
\end{equation}
Then $M_F$ is a positive, contractive, injective operator on $L^2(\mathbb R^+;\mathcal K)$, such that $\text{Ran}\,M_F$ is dense in $L^2(\mathbb R^+;\mathcal K)$ and $M_F$ is a Lyapunov operator in the
forward direction, i.e., for every $\psi\in L^2(\mathbb R^+;\mathcal K)$ we have
\begin{equation*}
 (\psi(t_2),M_F\,\psi(t_2))\leq (\psi(t_1),M_F\,\psi(t_1)),\qquad t_2\geq t_1\geq 0,\  \psi(t)=u_+(t)\psi,
\end{equation*}
and, moreover,
\begin{equation*}
 \lim_{t\to\infty} (\psi(t),M_F\,\psi(t))=0.
\end{equation*}
\hfill$\square$
\end{theorem}
\begin{theorem}
\label{zf_theorem}
Let $\Lambda_F:=M_F^{1/2}$. Then $\Lambda_F: L^2(\mathbb R^+;\mathcal K)\mapsto L^2(\mathbb R^+;\mathcal K)$ is a positive, contractive, injective operator such that $Ran\,\Lambda_F$ is dense in
$L^2(\mathbb R^+;\mathcal K)$. Furthermore, there exists a continuous, strongly contractive, one parameter semigroup $\{Z_F(t)\}_{t\in\mathbb R^+}$ with 
$Z_F(t): L^2(\mathbb R^+;\mathcal K)\mapsto L^2(\mathbb R^+;\mathcal K)$, such that for every $\psi\in L^2(\mathbb R^+;\mathcal K)$ we have
\begin{equation*}
 \Vert Z_F(t_2)\psi\Vert\leq\Vert Z_F(t_1)\psi\Vert,\quad t_2\geq t_1\geq 0
\end{equation*}
and
\begin{equation*}
 s-\lim_{t\to\infty} Z_F(t)=0
\end{equation*}
and the following intertwining relation holds:
\begin{equation}
\label{zf_intertwine}
 \Lambda_F u_+(t)=Z_F(t)\Lambda_F,\quad t\geq 0.
\end{equation}
\hfill$\square$
\end{theorem}
In a manner similar to the construction of a forward Lyapunov operator $M_F$ one is able to construct a backward Lyapunov operator $M_B$. The theorems analogous to Theorem \ref{mf_theorem} and 
Theorem \ref{zf_theorem} in this case are
\begin{theorem}
\label{mb_theorem}
Let $M_B: L^2(\mathbb R^+;\mathcal K)\mapsto L^2(\mathbb R^+;\mathcal K)$ be the operator defined by
\begin{equation}
\label{mb_definition}
 M_B:=(P_{\mathbb R^+}\hat P_- P_{\mathbb R^+})\big\vert_{L^2(\mathbb R^+;\mathcal K)}.
\end{equation}
Then $M_B$ is a positive, contractive, injective operator on $L^2(\mathbb R^+;\mathcal K)$, such that $Ran\,M_B$ is dense in $L^2(\mathbb R^+;\mathcal K)$ and $M_B$ is a Lyapunov operator in the backward
direction, i.e., for every $\psi\in L^2(\mathbb R^+;\mathcal K)$ we have
\begin{equation*}
 (\psi(t_2),M_B\,\psi(t_2))\leq (\psi(t_1), M_B\,\psi(t_1)),\quad t_2\leq t_1\leq 0,\ \ \psi(t)=u_+(t)\psi,
\end{equation*}
and, moreover,
\begin{equation*}
  \lim_{t\to -\infty}(\psi(t), M_B\,\psi(t))=0.
\end{equation*}
\hfill$\square$
\end{theorem}
\begin{theorem}
\label{zb_theorem}
Let $\Lambda_B:= M_B^{1/2}$. Then $\Lambda_B: L^2(\mathbb R^+;\mathcal K)\mapsto L^2(\mathbb R^+;\mathcal K)$ is a positive, contractive, injective operator such that $Ran\,\Lambda_B$ is dense in
$L^2(\mathbb R^+;\mathcal K)$. Furthermore, there exists a continuous, strongly contractive, one-parameter semigroup $\{Z_B(t)\}_{t\in\mathbb R^-}$ with 
$Z_B(t): L^2(\mathbb R^+;\mathcal K)\mapsto L^2(\mathbb R^+;\mathcal K)$, such that for every $\psi\in L^2(\mathbb R^+;\mathcal K)$ we have
\begin{equation*}
 \Vert Z_B(t_2)\psi\Vert\leq \Vert Z_B(t_1)\psi\Vert,\quad t_2\leq t_1\leq 0
\end{equation*}
and
\begin{equation*}
 s - \lim_{t\to -\infty} Z_B(t)=0
\end{equation*}
and the following intertwining relation holds:
\begin{equation}
\label{zb_intertwine}
 \Lambda_B u_+(t)= Z_B(t)\Lambda_B,\quad t\leq 0.
\end{equation}
\hfill$\square$
\end{theorem}
Note that Theorem \ref{mf_theorem} and Theorem \ref{mb_theorem} refer, respectively, to positive and negative times. However, due to the time translation
invariance of the evolution, the restriction that $t_2$, $t_1$ are non-negative in Theorem \ref{mf_theorem} and the $t_2$, $t_1$ are non-positive in Theorem
\ref{mb_theorem} can be removed (keeping the time ordering between $t_2$ and $t_1$ in both cases) and the Lyapunov property extends to all values of time. 
It is evident from Theorems 
\ref{mf_theorem} and \ref{mb_theorem} that both the forward and backward Lyapunov operators are defined on a rather abstract level in terms of certain functions spaces
and that no relation to any concrete class of physical problems has been made yet. We amend this by introducing Lyapunov operators specifically in the context of quantum mechanical scattering problems.
\par In the following we consider quantum mechanical scattering problems satisfying the following two assumptions:
\begin{description}
\item{(i)} Let $\mathcal H$ be a separable Hilbert space corresponding to a given quantum mechanical scattering problem. Assume that a self-adjoint "free" unperturbed Hamiltonian $H_0$ and a self-adjoint 
perturbed Hamiltonian $H$ are defined on $\mathcal H$ and form a complete scattering system, i.e., we assume that the M\o ller wave operators $\Omega_\pm\equiv\Omega_\pm(H_0,H)$ exist and are complete.
\item{(ii)} We assume that $\sigma_{ac}(H)=\sigma_{ac}(H_0)=\mathbb R^+$. Moreover, we assume that the multiplicity of the absolutely continuous 
spectrum is uniform over $\mathbb R^+$. 
\end{description}
Under assumptions (i)-(ii) above, there exist two mappings $\hat W^{\text{\tiny QM}}_\pm:\mathcal H_{ac}\mapsto L^2(\mathbb R^+;\mathcal K)$ that map the subspace 
$\mathcal H_{ac}\subseteq\mathcal H$ isometrically onto the function space $L^2(\mathbb R^+;\mathcal K)$ for some Hilbert space $\mathcal K$ whose dimension corresponds to the multiplicity of 
$\sigma_{ac}(H)$ and the Schr\"odinger evolution $U(t)=\exp(-iHt)$ is represented by the group
$\{u_+(t)\}_{t\in\mathbb R}$. The represnetation of the scattering problem in the function space $L^2(\mathbb R^+;\mathcal K)$ obtained by applying 
$\hat W^{\text{\tiny QM}}_+$ is known as the outgoing energy 
representation and is a spectral representation for $H$ in which the action of $H$ is represented by multiplication by the independent variable. In a similar manner, the representation obtain by applying 
the mapping $\hat W^{\text{\tiny QM}}_-$ is another spectral representation for $H$, known as the incoming energy representation of the problem. We note that all of the objects $M_F$, $M_B$, $\Lambda_F$, 
$\Lambda_B$, $Z_F(t)$ and $Z_B(t)$ in Theorems \ref{mf_theorem}-\ref{zb_theorem} are defined on the level of such spectral representations of $H$ and their construction is made irrespective of the specific spectral representation in which one is working. However, when applied to scattering problems, we need to distinguish between the corresponding objects defined within the incoming energy representation and the outgoing energy representation.
\par The mappings $\hat W^{\text{\tiny QM}}_+$ and $\hat W^{\text{\tiny QM}}_-$ correspond, respectively, to incoming and outgoing solutions of the 
the Lippmann-Schwinger equation. If $\{\phi^-_{E,\xi}\}_{E\in\mathbb R^+,\,\xi\in\Xi}$ are outgoing solutions of the Lippmann-Schwinger equation, where $\xi$ corresponds
to degeneracy indices of the energy $E$, and if $\{\phi^+_{E,\xi}\}_{E\in\mathbb R^+,\,\xi\in\Xi}$ are incoming solutions of the Lippmann-Schwinger equation, and $\psi\in\mathcal H_{ac}$ is any scattering state, then
\begin{eqnarray*}
 (\hat W^{\text{\tiny QM}}_+\psi)(E,\xi)&=&(\phi^-_{E,\xi},\psi)\\
 (\hat W^{\text{\tiny QM}}_-\psi)(E,\xi)&=&(\phi^+_{E,\xi},\psi).\\
\end{eqnarray*}
With the help of the two mapping $\hat W^{\text{\tiny QM}}_\pm$ which are, in fact, associated with the two M\o ller wave operators $\Omega_\pm$, we define the 
forward Lyapunov operator for the quantum scattering problem to be
\begin{equation}
\label{QM_forward_lyapunov_op_construction}
 M_+:=\left(\hat W^{\text{\tiny QM}}_+\right)^{-1}M_F \hat W^{\text{\tiny QM}}_+.
\end{equation}
By Theorem \ref{mf_theorem} the operator $M_+$ is a positive, contractive, injective operator on $\mathcal H_{ac}$, such that $\text{Ran}\,M_+$ is dense in $\mathcal H_{ac}$ and $M_+$ is a forward 
Lyapunov operator with respect to the quantum evolution on $\mathcal H_{ac}$. Similarly, the backward Lyapunov operator for the quantum scattering problem is defined to be
\begin{equation}
\label{QM_backward_lyapunov_op_construction}
 M_-:=\left(\hat W^{\text{\tiny QM}}_-\right)^{-1}M_B \hat W^{\text{\tiny QM}}_-.
\end{equation}
According to Theorem \ref{mb_theorem} the operator $M_-$ is a positive, contractive, injective operator on $\mathcal H_{ac}$, such that $\text{Ran}\,M_-$ is dense in $\mathcal H_{ac}$ and $M_-$ is a backward
Lyapunov operator with respect to the quantum evolution on $\mathcal H_{ac}$. (the Lyapunov operators $M_+$ and $M_-$ are identical, respectively, to the outgoing forward Lyapunov operator $M_{F,+}$ 
and the incoming backward Lyapunov operator $M_{B,-}$ of Ref. \cite{S2}). Eqns. (\ref{QM_forward_lyapunov_op_construction}) and (\ref{QM_backward_lyapunov_op_construction}) are to be compared to 
Eqns. (\ref{LP_forward_lyapunov_op_construction}) and (\ref{LP_backward_lyapunov_op_construction}). 
\par We presently show that the Lax-Phillips Lyapunov operators $P_\pm$ and the quantum mechanical Lyapunov operators $M_\pm$ are but two particular instances of a more general construction. 
Consider a scattering problem, defined on a Hilbert space $\mathcal H$, for which $H$ is the generator of the evolution 
group $\{U(t)\}_{t\in\mathbb R}$ of the system under consideration, i.e., $U(t)=\exp(-iHt)$. Denote by $\sigma_{ac}(H)$ the absolutely continuous part of the spectrum of $H$ and let 
$\mathcal H_{ac}\subseteq\mathcal H$ denote the subspace corresponding to $\sigma_{ac}(H)$. Assume, furthermore, that the multiplicity of the absolutely continuous spectrum is uniform over 
$\sigma_{ac}(H)$ and that the M\o ller wave operators exist and are complete. Then there exist two unitary mappings $\hat W_\pm: \mathcal H_{ac}\mapsto L^2(\mathbb R;\mathcal K)$, corresponding to the outgoing and incoming wave operators of the problem, which map 
$\mathcal H_{ac}$ onto a function space $L^2(\sigma_{ac}(H),\mathcal K)\subseteq L^2(\mathbb R,\mathcal K)$, where $\mathcal K$ is a Hilbert space whose dimension corresponds to the multiplicity of 
$\sigma_{ac}(H)$ and such that $\hat W_\pm$ maps the action of the generator of evolution $H$ into multiplication by the independent variable in $L^2(\sigma_{ac}(H),\mathcal K)$. The two representations of the 
scattering problem thus obtained are the incoming and outgoing energy representations for the problem.
Let $P_{\sigma_{ac}(H)}:L^2(\mathbb R,\mathcal K)\mapsto L^2(\mathbb R,\mathcal K)$ be the orthogonal projection in $L^2(\mathbb R,\mathcal K)$ on the subspace $L^2(\sigma_{ac}(H),\mathcal K)$. Hence,
If $P_{ac}(H):\mathcal H\mapsto\mathcal H$ is the orthogonal projection in $\mathcal H$ on $\mathcal H_{ac}$, we have
\begin{equation*}
  \hat W_\pm P_{ac}(H)\mathcal H =\hat W_\pm\mathcal H_{ac}=L^2(\sigma_{ac}(H),\mathcal K)=P_{\sigma_{ac}(H)}L^2(\mathbb R,\mathcal K).
\end{equation*}
Define two operators $M_{\pm}(H)\,:\,\mathcal H_{ac}\mapsto\mathcal H_{ac}$ by
\begin{equation}
\label{general_M_pm_definition}
 M_\pm(H):=\hat W_\pm^{-1}P_{\sigma_{ac}(H)}\hat P_\pm P_{\sigma_{ac}(H)}\hat W_\pm,
\end{equation}
where $\hat P_\pm$ are, respectively, the projections in $L^2(\mathbb R,\mathcal K)$ on the Hardy subspaces $\mathcal H^2_\pm(\mathbb R,\mathcal K)$. It is readily verified that $M_\pm(H)$ are positive, 
contractive, operators on $\mathcal H_{ac}$. Now, if $\sigma_{ac}(H)=\mathbb R^+$ we find that $P_{\sigma_{ac}(H)}=P_{\mathbb R_+}$ and hence in this case we obtain
\begin{equation*}
 M_+(H)=\hat W_+^{-1}P_{\sigma_{ac}(H)}\hat P_+ P_{\sigma_{ac}(H)}\hat W_+ = \hat W_+^{-1}P_{\mathbb R^+}\hat P_+ P_{\mathbb R^+}\hat W_+ 
 = \hat W_+^{-1}M_F \hat W_+ = M_+, 
\end{equation*}
and similarly
\begin{equation*}
 M_-(H)=\hat W_-^{-1}P_{\sigma_{ac}(H)}\hat P_- P_{\sigma_{ac}(H)}\hat W_- = \hat W_-^{-1}P_{\mathbb R^+}\hat P_- P_{\mathbb R^+}\hat W_- 
 = \hat W_-^{-1}M_B \hat W_- = M_-. 
\end{equation*}
If, on the other hand, we consider the Lax-Phillips scattering theory the generator $K$ of the unitary evolution group $\{U(t)\}_{t\in\mathbb R}=\{e^{-iKt}\}_{t\in\mathbb R}$ defined on the Lax-Phillips
Hilbert space $\mathcal H^{\text{\tiny LP}}$, satisfies $\sigma_{ac}(K)=\mathbb R$ and we have $P_{\sigma_{ac}(K)}=I_{L^2(\mathbb R,\mathcal K)}$. Therefore, in this case we have
\begin{equation*}
 M_+(K)=\hat W_+^{-1}P_{\sigma_{ac}(K)}\hat P_+ P_{\sigma_{ac}(K)}\hat W_+ 
 = \hat W_+^{-1}I_{L^2(\mathbb R,\mathcal K)}\hat P_+ I_{L^2(\mathbb R,\mathcal K)}\hat W_+ = \hat W_+^{-1}\hat P_+ \hat W_+ = P_+, 
\end{equation*}
and
\begin{equation*}
 M_-(K)=\hat W_-^{-1} P_{\sigma_{ac}(K)}\hat P_- P_{\sigma_{ac}(K)}\hat W_- 
 = \hat W_-^{-1}I_{L^2(\mathbb R,\mathcal K)}\hat P_- I_{L^2(\mathbb R,\mathcal K)}\hat W_- = \hat W_-^{-1}\hat P_- \hat W_- = P_-.
\end{equation*}
\par Since the operators $M_\pm(H)$ defined in Eq. (\ref{general_M_pm_definition}) are positive, contractive operators on $\mathcal H_{ac}$ then their square roots
\begin{equation*}
 \Lambda_\pm(H):=M^{1/2}_\pm(H)
\end{equation*}
are well defined as operators on $\mathcal H_{ac}$. Define a family of operators $\{Z_{app}(t)\}_{t\geq 0}\,:\,\mathcal H_{ac}\mapsto\mathcal H_{ac}$ by
\begin{equation}
\label{general_approx_lp_semigroup_def}
 Z_{app}(t):=\Lambda_+(H)U(t)\Lambda_-(H),\quad t\geq 0.
\end{equation}
Applying the definitions of $\Lambda_\pm(H)$ in the Lax-Phillips scattering theory we obtain 
\begin{equation*}
 \Lambda_\pm(K):=M^{1/2}_\pm(K)=P_\pm^{1/2}=P_\pm.
\end{equation*}
Eq. (\ref{general_approx_lp_semigroup_def}) then yields 
\begin{equation*}
 Z_{app}(t)=\Lambda_+(K)U(t)\Lambda_-(K)=P_+U(t)P_- = Z_{\text{\tiny LP}}(t),\quad t\geq 0,
\end{equation*}
so that $\{Z_{app}(t)\}_{t\geq 0}$ is in this case exactly the Lax-Phillips semigroup $\{Z_{\text{\tiny LP}}(t)\}_{t\geq 0}$. Applying the definitions in the case of a quantum 
mechanical scattering problem satisfying assumptions (i)-(ii) above we get
\begin{equation*}
 \Lambda_\pm(H):=M^{1/2}_\pm(H)=M_\pm^{1/2}=\Lambda_\pm,
\end{equation*}
where $\Lambda_\pm:=M_\pm^{1/2}$, and hence
\begin{equation*}
 Z_{app}(t)=\Lambda_+(H)U(t)\Lambda_-(H)=\Lambda_+U(t)\Lambda_-,\quad t\geq 0.
\end{equation*}
The family of operators $\{Z_{app}(t)\}_{t\geq 0}$ is identified in this case as the approximate Lax-Phillips semigroup of Eq. (\ref{approx_lax_phillips_semigroup}). 
\par We make a formal definition of the approximate Lax-Phillips semigroup for a scattering problem satisfying conditions (i)-(ii):
\begin{definition}[Approximate Lax-Phillips semigroup]
Consider scattering problem satisfying assumptions (i)-(ii) above. Let $\Lambda_+=M_+^{1/2}$ and $\Lambda_-=M_-^{1/2}$ where
$M_+$ is the forward Lyapunov operator and $M_-$ is the backward Lyapunov operator for the problem. Then the approximate Lax-Phillips
semigroup is defined to be the family of operators $\{Z_{app}(t)\}_{t\in\mathbb R^+}$, $Z_{app}(t): \mathcal H_{ac}\mapsto\mathcal H_{ac}$ defined by 
\begin{equation}
\label{approx_lp_semigroup_def}
 Z_{app}(t):=\Lambda_+U(t)\Lambda_-,\quad t\geq 0,\ \ U(t)=e^{-iHt}.
\end{equation}
\hfill$\square$
\end{definition}
Note that Theorem \ref{zf_theorem} implies
that there exists a continuous, strongly contractive, one parameter semigroup $\{Z_+(t)\}_{t\in\mathbb R^+}$ with 
$Z_+(t): \mathcal H_{ac}\mapsto\mathcal H_{ac}$, such that for every $\psi\in \mathcal H_{ac}$ we have
\begin{equation*}
 \Vert Z_+(t_2)\psi\Vert\leq\Vert Z_+(t_1)\psi\Vert,\quad t_2\geq t_1\geq 0
\end{equation*}
and
\begin{equation*}
 s-\lim_{t\to\infty} Z_+(t)=0
\end{equation*}
and the following intertwining relation holds
\begin{equation}
\label{z_+_intertwine}
 \Lambda_+ U(t)=Z_+(t)\Lambda_+,\quad t\geq 0.
\end{equation}
In fact, we have $Z_+(t)=(\hat W_+^{\text{\tiny QM}})^{-1}Z_F(t)\hat W_+^{\text{\tiny QM}}$, $t\geq 0$, where $Z_F(t)$ are elements of the semigroup $\{Z_F(t)\}_{t\geq 0}$ in Theorem \ref{zf_theorem}. Similarly, Theorem \ref{zb_theorem} implies
that there exists a continuous, strongly contractive, one parameter semigroup $\{Z_-(t)\}_{t\in\mathbb R^-}$ with $Z_-(t): \mathcal H_{ac}\mapsto\mathcal H_{ac}$, such that for every 
$\psi\in \mathcal H_{ac}$ we have
\begin{equation*}
 \Vert Z_-(t_2)\psi\Vert\leq\Vert Z_-(t_1)\psi\Vert,\quad t_2\leq t_1\leq 0
\end{equation*}
and
\begin{equation*}
 s-\lim_{t\to -\infty} Z_-(t)=0
\end{equation*}
and the following intertwining relation holds
\begin{equation}
\label{z_-_intertwine}
 \Lambda_- U(t)=Z_-(t)\Lambda_-,\quad t\leq 0.
\end{equation}
It is precisely due to the central importance of the intertwining relations in Eqns. (\ref{z_+_intertwine}) and (\ref{z_-_intertwine}) that the definition of the approximate Lax-Phillips semigroup in 
Eqns. (\ref{general_approx_lp_semigroup_def}) and (\ref{approx_lp_semigroup_def}) is made using the square roots $\Lambda_\pm$ of the Lyapunov operators $M_\pm$ and not the Lyapunov operators themselves 
(as mentioned in the introduction, it is evident that in the Lax-Phillips case a definition using the Lyapunov operators or their square roots would yield the same family of objects).
\par We complete the set of relations between constructions of the Lax-Phillips theory and the corresponding constructions for quantum mechanical scattering by considering the incoming and outgoing 
representations. Observe that the outgoing (spectral or translation) representations of the Lax-Phillips theory are distinguished by the representation of the outgoing subspace $\mathcal D_+$. Hence, for 
example, if $\hat W^{\text{\tiny LP}}_+$ is the mapping of $\mathcal H^{\text{\tiny LP}}$ onto the outgoing spectral representation then 
$\hat W^{\text{\tiny LP}}_+\mathcal D_+ = \mathcal H^2_-(\mathbb R;\mathcal K)$ and $\hat W^{\text{\tiny LP}}_+(\mathcal K\oplus\mathcal D_-)=\mathcal H^2_+(\mathbb R;\mathcal K)$. Thus the outgoing 
representations are centered on the separation of the outgoing part of an evolving  state $\psi(t)=U(t)\psi$ from the other components of that state which is achieved by the application of the projection 
$P_+$. This implies a decomposition of an evolving state $\psi(t)=\exp(-iKt)\psi$, corresponding to an initial state $\psi\in\mathcal H^{\text{\tiny LP}}$, according to
\begin{equation}
\label{LP_forward_transition_rep}
 \psi(t)=P_+\psi(t)+P_+^\perp\psi(t)=\psi^b_+(t)+\psi^f_+(t),
\end{equation}
where $\psi^b_+(t):=P_+\psi(t)$ and $\psi^f_+(t):=P_+^\perp\psi(t)$. It is readily verified, using the outgoing translation representation, that $\psi^b_+(t)$ is a \emph{backward asymptotic component} of 
$\psi(t)$, i.e., $\psi_+^b(t)$ vanishes in the forward time asymptote as $t\to\infty$ and is asymptotic to $\psi(t)$ in the backward time asymptote as $t\to-\infty$. Similarly, $\psi^f_+(t)$ is a 
\emph{forward asymptotic component} of $\psi(t)$, i.e., $\psi^f_+(t)$ vanishes in the backward time asymptote as $t\to -\infty$ and is asymptotic to $\psi(t)$ in the forward time asymptote as $t\to\infty$. 
The evolution of $\psi(t)$ is then represented as a transition from $\psi^b_+(t)$ to $\psi^f_+(t)$. We call the representation of the evolution obtained by the decomposition in Eq. (\ref
{LP_forward_transition_rep}) a \emph{forward transition representation} and emphasize again its direct relation to the outgoing (translation or spectral) representations in the Lax-Phillips theory. Note 
that the name given to this representation of the evolution registers both the fact that the representation involves a transition between different 
components of the evolving state and the fact that the decomposition in Eq. (\ref{LP_forward_transition_rep}) is obtained using the forward Lyapunov operator $P_+$.
\par Following a similar line of argument we may use the backward Lyapunov operator, i.e., the projection $P_-$, to obtain a decomposition of an evolving state $\psi(t)$ in the form
\begin{equation}
\label{LP_backward_transition_rep}
 \psi(t)=P_-\psi(t)+P_-^\perp\psi(t)=\psi^b_-(t)+\psi^f_-(t),
\end{equation}
where $\psi^f_-(t):=P_-\psi(t)$ and $\psi^b_-(t):=P_-^\perp\psi(t)=(I-P_-)\psi(t)$. Here $\psi^f_-(t)$ is a forward asymptotic component of $\psi(t)$ and $\psi^b_-(t)$ is a backward asymptotic component of 
$\psi(t)$ and we obtain another transition representation of the evolution of $\psi(t)$ which we call the \emph{backward transition representation}. In a manner similar to the case of the forward transition 
representation, the backward transition representation is directly associated with the Lax-Phillips incoming (spectral or translation) representations. It is evident from the structure of the Lax-Phillips 
theory that such transition representations are useful for the description of transient phenomena in scattering processes, such as resonances.
\par Turning to the quantum mechanical case we may define forward and backward transition representations analogous to those defined in the Lax-Phillips theory using the following two propositions,  
proved in Ref. \cite{S2}:
\begin{proposition}
\label{forward_trans_rep}
For $\psi(t)=u_+(t)\psi$, $\psi\in L^2(\mathbb R_+;\mathcal K)$, $t\in\mathbb R$, apply the following decomposition
\begin{equation}
\label{forward_trans_decomp}
 \psi(t)=\psi_F^b(t)+\psi_F^f(t)
\end{equation}
where
\begin{equation*}
 \psi_F^b(t):=\Lambda_F\psi(t),\qquad \psi_F^f(t):=(I-\Lambda_F)\psi(t).
\end{equation*}
Then
\begin{eqnarray*}
 \lim_{t\to -\infty}\Vert\psi(t)-\psi_F^b(t)\Vert=0&,&\quad \lim_{t\to\infty}\Vert\psi_F^b(t)\Vert=0,\\
 \lim_{t\to -\infty}\Vert\psi_F^f(t)\Vert=0&,&\quad \lim_{t\to\infty}\Vert\psi(t)-\psi_F^f(t)\Vert=0.
\end{eqnarray*}
\hfill$\square$
\end{proposition}
and
\begin{proposition}
\label{backward_trans_rep}
For $\psi(t)=u_+(t)\psi$, $\psi\in L^2(\mathbb R_+;\mathcal K)$, $t\in\mathbb R$, apply the following decomposition
\begin{equation}
\label{backward_trans_decomp}
 \psi(t)=\psi_B^b(t)+\psi_B^f(t)
\end{equation}
where
\begin{equation*}
 \psi_B^b(t):=(I-\Lambda_B)\psi(t),\qquad \psi_B^f(t):=\Lambda_B\psi(t).
\end{equation*}
Then
\begin{eqnarray*}
 \lim_{t\to -\infty}\Vert\psi(t)-\psi_B^b(t)\Vert=0&,&\quad \lim_{t\to\infty}\Vert\psi_B^b(t)\Vert=0,\\
 \lim_{t\to -\infty}\Vert\psi_B^f(t)\Vert=0&,&\quad \lim_{t\to\infty}\Vert\psi(t)-\psi_B^f(t)\Vert=0.
\end{eqnarray*}
\hfill$\square$
\end{proposition}
Proposition \ref{forward_trans_rep} states that $\psi(t)$ can be decomposed into a sum of two components, $\psi_F^b(t)$ and $\psi_F^f(t)$ such that 
$\psi_F^b(t)$ is a backward asymptotic component and $\psi_F^f(t)$ is a forward asymptotic component of $\psi(t)$. Via the decomposition in Eq. (\ref{forward_trans_decomp}) the evolution of 
$\psi(t)$ is represented as a transition from the backward asymptotic component to the forward asymptotic component and we obtain a transition representation of the evolution which, by the fact that the 
decomposition is defined using the (square root of the) forward Lyapunov operator, is a forward transition representation. 
By Proposition \ref{backward_trans_rep} we have a different decomposition of $\psi(t)$ into a backward asymptotic component $\psi_F^b(t)$ and a forward asymptotic component $\psi_F^f(t)$. The 
resulting transition representation of the evolution of $\psi(t)$ in this case is a backward transition representation, i.e., the decomposition of the evolving state $\psi(t)$ into the two components in 
Eq. (\ref{backward_trans_decomp}) is achieved via the use of the backward Lyapunov operator.
\par Consider a scattering problem satisfying assumptions (i)-(ii) above. Defining the forward Lyapunov operator $M_+$ for the scattering problem as in 
Eq. (\ref{QM_forward_lyapunov_op_construction}) and 
noting that $\Lambda_+=(\hat W^{\text{\tiny QM}}_+)^{-1}M_F^{1/2} \hat W^{\text{\tiny QM}}_+$ we immediately obtain, using proposition \ref{forward_trans_rep}, a forward transition representation for the quantum 
evolution. The formal definition of this representation is:
\begin{definition}[forward transition representation]
\label{outgoing_forward_trans_rep_def}
Let $\Lambda_+:=M_+^{1/2}$ be the square root of $M_+$. For any $\psi\in\mathcal H_{ac}$ the forward transition representation of the evolution of $\psi$ is defined to be the decomposition
\begin{equation*}
 \psi(t)=\psi_+^b(t)+\psi_+^f(t),
\end{equation*}
where $\psi_+^b(t):=\Lambda_+\psi(t)$, $\psi_+^f(t):=(I-\Lambda_+)\psi(t)$, $\psi(t)=U(t)\psi$ and $U(t)=\exp(-iHt)$ is the Schr\"odinger evolution in $\mathcal H$.\hfill $\square$ 
\end{definition}
The asymptotic behavior over time of the two components $\psi_+^b(t)$, $\psi_+^f(t)$ of $\psi(t)$ follows directly from Proposition \ref{forward_trans_rep}. 
The backward transition representation is defined in a similar manner following the definition of the backward Lyapunov operator $M_-$ for the scattering problem 
in Eq. (\ref{QM_backward_lyapunov_op_construction}) and the fact that $\Lambda_- =(\hat W^{\text{\tiny QM}}_-)^{-1}M_B^{1/2} \hat W^{\text{\tiny QM}}_-$:
\begin{definition}[backward transition representation]
\label{incoming_backward_trans_rep_def}
\par Let $\Lambda_-:=M_-^{1/2}$ be the square root of $M_-$.
For any $\psi\in\mathcal H_{ac}$ the backward transition representation of the evolution of $\psi$ is defined to be the decomposition
\begin{equation*}
 \psi(t)=\psi_-^b(t)+\psi_-^f(t),
\end{equation*}
where $\psi_-^b(t):=(I-\Lambda_-)\psi(t)$, $\psi_-^f(t):=\Lambda_-\psi(t)$, $\psi(t)=U(t)\psi$ and $U(t)=\exp(-iHt)$ is the Schr\"odinger evolution in 
$\mathcal H$.\hfill $\square$ 
\end{definition}
Of course, the asymptotic behavior over time of the two components $\psi_-^b(t)$, $\psi_-^f(t)$ of the backward transition representation follows directly from Proposition 
\ref{backward_trans_rep}. 
\par By defining the two transition representations in Definitions \ref{outgoing_forward_trans_rep_def} and \ref{incoming_backward_trans_rep_def} we complete the construction within the framework of 
quantum mechanics of objects and representations analogous to the central objects and representations of the Lax-Phillips theory. We may then extend Eq. (\ref{lp_qm_analogy_ver_1}) as follows:
\begin{eqnarray}
\label{lp_qm_analogy_ver_2}
 \underline{\text{LP scattering theory}} &{ }&  \underline{\text{QM scattering theory}}\notag\\
 U(t)=e^{-iKt} &\Longleftrightarrow& U(t)=e^{-iHt}\notag\\
 P_\pm &\Longleftrightarrow& \Lambda_\pm\\
 \psi(t)=P_+\psi(t)+P_+^\perp\psi(t) &\Longleftrightarrow& \psi(t)=\Lambda_+\psi(t)+(I-\Lambda_+)\psi(t)\notag\\
 \psi(t)=P_-^\perp\psi(t)+P_-\psi(t) &\Longleftrightarrow& \psi(t)=(I-\Lambda_-)\psi(t)+\Lambda_-\psi(t)\notag\\
 Z_{\text{\tiny{LP}}}(t)=P_+U(t)P_-,\ t\geq 0 &\Longleftrightarrow& Z_{app}(t)=\Lambda_+U(t)\Lambda_-,\ t\geq 0\notag\\
 \hat S_{\text{\tiny LP}}(E),\ \ E\in\mathbb R &\Longleftrightarrow& \hat S_{\text{\tiny QM}}(E),\ \  E\in\mathbb R^+\notag
\end{eqnarray}
\par We remark that the forward transition representation, corresponding to the decomposition on the right hand side of the third line in Eq. (\ref{lp_qm_analogy_ver_2}) above, has already been 
used successfully in Ref. \cite{S2} for the description of quantum mechanical resonance scattering processes and results in a clear separation of the outgoing probability waves from the incoming waves 
and the formation of a resonance (the forward transition representation is called in Ref. \cite{S2} the outgoing forward transition representation). 
\section{Resonance poles, resonance states and the approximate Lax-Phillips semigroup in quantum mechanical scattering}
\label{res_poles_and_states_and_approx_LP_semigroup}
\par Upon completion of the set of relations in Eq. (\ref{lp_qm_analogy_ver_2}) we are left with an important task, i.e., to establish  in the context of quantum mechanical scattering a theorem analogous 
to Theorem \ref{LP_main_result} relating resonance poles of the Lax-Phillips scattering matrix to eigenvalues and eigenvectors of the Lax-Phillips semigroup. Hence, an appropriate definition of resonance 
states and investigation of their relation to the approximate Lax-Phillips semigroup, defined in the previous section, is a central ingredient in the development of the formalism introduced in the present 
work. Of course, we do not expect to define resonance states as exact eigenvectors of elements $Z_{app}(t)$ of the approximate Lax-Phillips semigroup since, as its name suggests, it is not an exact 
semigroup. However, we may try to find resonance states, associated with resonance poles of the quantum mechanical scattering matrix, which are eigenvectors of $Z_{app}(t)$ in some approximate sense and 
make an effort to quantify the quality of such an approximation. 
\par The problem of the definition of appropriate resonance states corresponding to resonance poles of the scattering matrix in quantum mechanical scattering has been addressed in the context of the recent 
development of the formalism of \emph{semigroup decomposition of resonance evolution} \cite{S3,S4,SHV} (of course, there are several other formalisms for dealing with the problem of scattering resonances 
in quantum mechanics, notably complex scaling \cite{AC,BC,Sim1,Sim2,Hun,SZ,HS} and rigged Hilbert spaces \cite{BaSch,Baum,BG,HoSi,PGS}. Here we consider the framework most suitable, in terms of its 
mathematical constructions, for the development of the formalism introduced in the present paper). The semigroup decomposition formalism utilizes basic mathematical constructions of the Lax-Phillips 
and the Sz.-Nagy-Foias theory for the formulation of a time dependent theory for the description of the evolution of scattering resonances in quantum mechanics. Significant progress has been achieved in the 
development of the structure of the general formalism under certain simplifying assumptions which we continue to apply in the present paper. Thus, we add to assumptions (i)-(ii) above the assumptions
\begin{description}
\item{(iii)} The absolutely continuous spectrum of the free Hamiltonian $H_0$ and the full Hamiltonian $H$ is simple, i.e., the multiplicity of the absolutely continuous spectrum is one.
\item{(iv)} Denote by $\mathbb C^+$, respectively by $\mathbb C^-$ the upper and lower half-planes of the complex plane $\mathbb C$. We assume that the S-matrix in the energy representation, denoted by 
$\hat S_{\text{\tiny QM}}(\cdot)$, has an extension into a function 
$\hat{\mathcal S}_{\text{\tiny QM}}(\cdot)$, holomorphic in some region $\Sigma^+\subset\mathbb C^+$ above the positive real axis 
$\mathbb R^+$ and having an analytic continuation across $\mathbb R^+$ into a region $\Sigma^-\subset\mathbb C^-$ such that the resulting analytically continued function, again denoted by 
$\hat{\mathcal S}_{\text{\tiny QM}}(\cdot)$, is meromorphic in an open, simply connected region $\Sigma=\Sigma^+\cup\Sigma^-\cup(\Sigma\cap\mathbb R^+)$. We assume that $\hat{\mathcal S}_{\text{\tiny QM}}(\cdot)$ has a single, simple, resonance pole at a point $z=\mu\in\Sigma^-$ and no other singularity in $\overline\Sigma$ ($\overline\Sigma$ is the closure of $\Sigma$).
\end{description}
We emphasize that the semigroup decomposition formalism may be applied under much less stringent conditions than those assumed here. However, conditions (iii)-(iv) make the discussion below more transparent and, in fact, facilitate the development of the formalism in the present paper.
\par It is shown in Ref. \cite{S2} that if we apply the operator $\Lambda_+$ to any state $\psi\in\mathcal H_{ac}$ then the resonance pole of the S-matrix $\hat{\mathcal S}_{\text{\tiny QM}}(\cdot)$ at $z=\mu$ (see assumption (iv) above) induces a decomposition of the state $\psi_{\Lambda_+}:=\Lambda_+\psi$ of the form
\begin{equation}
\label{psi_lambda_plus_decomp}
 \psi_{\Lambda_+}=\Lambda_+\psi=b(\psi;\mu)+(\psi_\mu^{app},\psi)\Vert\psi_\mu^{res}\Vert^{-2}\psi_\mu^{res}.
\end{equation}
The state $\psi^{app}_\mu\in\mathcal H_{ac}$ is referred to as an \emph{approximate resonance state} and the state $\psi_\mu^{res}\in\mathcal H_{ac}$ is referred to as the \emph{resonance state} 
corresponding to the resonance pole at $z=\mu$. Note that in Ref. \cite{S2} the resonance state $\psi_\mu^{res}$ is denoted by $\psi_\mu^{ir}$. The states $\psi_\mu^{app}$ and $\psi_\mu^{res}$
are related. In fact, we have
\begin{equation*}
 \psi_\mu^{app}=\Lambda_+\psi_\mu^{res}.
\end{equation*}
Thus we may write equation (\ref{psi_lambda_plus_decomp}) in the form
\begin{multline}
\label{psi_lambda_plus_decomp_ver_2}
 \psi_{\Lambda_+}=\Lambda_+\psi=b(\psi;\mu)+(\psi_\mu^{app},\psi)\Vert\psi_\mu^{res}\Vert^{-2}\psi_\mu^{res}
 =b(\psi;\mu)+(\Lambda_+\psi_\mu^{res},\psi)\Vert\psi_\mu^{res}\Vert^{-2}\psi_\mu^{res}=\\
 =b(\psi;\mu)+(\psi_\mu^{res},\Lambda_+\psi)\Vert\psi_\mu^{res}\Vert^{-2}\psi_\mu^{res}
 =b(\psi;\mu)+(\psi_\mu^{res},\psi_{\Lambda_+})\Vert\psi_\mu^{res}\Vert^{-2}\psi_\mu^{res}=\\
 =b(\psi;\mu)+P_{res}\psi_{\Lambda_+},
\end{multline}
where $P_{res}$ is the projection on the subspace $\mathcal H_{res}:=P_{res}\mathcal H_{ac}\subset\mathcal H_{ac}$ spanned by the resonance state $\psi_\mu^{res}$. We emphasize again that the decomposition 
in Eq. (\ref{psi_lambda_plus_decomp}) or Eq. (\ref{psi_lambda_plus_decomp_ver_2}) is not arbitrary but naturally induced by the existence of the resonance pole 
of the scattering matrix $\hat{\mathcal S}_{\text{\tiny QM}}(\cdot)$. It is shown furthermore in Ref. \cite{S2} that 
\begin{equation}
\label{z_+_eigenvector}
 Z_+(t)\psi_\mu^{res}=e^{-i\mu t}\psi_\mu^{res},\quad t\geq 0.
\end{equation}
where $Z_+(t)$ are elements of the semigroup $\{Z_+(t)\}_{t\geq 0}$, appearing on the right hand side of Eq. (\ref{z_+_intertwine}).
Now define a linear subspace $(\mathcal H_{ac})_{\Lambda_+}:=\Lambda_+\mathcal H_{ac}$. Since $\Lambda_+$ is injective and since $\text{Ran}\,\Lambda_+$ is 
dense in $\mathcal H_{ac}$ we have that $(\mathcal H_{ac})_{\Lambda_+}$ is a dense linear subspace of $\mathcal H_{ac}$ and for any state 
$\varphi\in(\mathcal H_{ac})_{\Lambda_+}$ the state $\tilde\varphi=\Lambda_+^{-1}\varphi$ is well defined in $\mathcal H_{ac}$.
Taking arbitrary states $\varphi\in(\mathcal H_{ac})_{\Lambda_+}$, $\psi\in\mathcal H_{ac}$ we may use Eqns. (\ref{psi_lambda_plus_decomp}), 
(\ref{z_+_intertwine}) and (\ref{z_+_eigenvector}) to obtain
\begin{multline}
\label{abstract_semigroup_decomp}
 (\varphi,U(t)\psi)=(\Lambda_+\tilde\varphi,U(t)\psi)=(\tilde\varphi,\Lambda_+U(t)\psi)
 =(\tilde\varphi,Z_+(t)\Lambda_+\psi)=(\tilde\varphi,Z_+(t)\psi_{\Lambda_+})=\\
 =(\tilde\varphi,Z_+(t)[b(\psi;\mu)+P_{res}\psi_{\Lambda_+}])=B(\varphi,\psi,\mu,t)+(\tilde\varphi,Z_{\text{\tiny QM}}(t)\psi_{\Lambda_+}),\quad t\geq 0,
\end{multline}
where $B(\varphi,\psi,\mu,t):=(\tilde\varphi,Z_+(t)b(\psi;\mu))$ and $Z_{\text{\tiny QM}}(t):=Z_+(t)P_{res}$. Since $P_{res}$ is a projection on a subspace of eigenvectors of $Z_+(t)$ the family of operators
$\{Z_{\text{\tiny QM}}(t)\}_{t\geq 0}$ annihilates $\mathcal H_{ac}\backslash\mathcal H_{res}$ and forms a continuous, one parameter, contractive semigroup on the subspace $\mathcal H_{res}$. 
The right hand side of Eq. (\ref{abstract_semigroup_decomp}) is the semigroup decomposition of the matrix element $(\varphi,U(t)\psi)$ of the Schr\"odinger evolution for $t\geq 0$.
\par Using the fact that, for $t\geq 0$, we have $Z_{\text{\tiny QM}}(t)=Z_+(t)P_{res}=e^{-i\mu t}P_{res}$ we obtain from Eq. (\ref{abstract_semigroup_decomp}) 
\begin{equation*}
 (\varphi,U(t)\psi)=B(\varphi,\psi,\mu,t)+(\tilde\varphi,Z_{\text{\tiny QM}}(t)\psi_{\Lambda_+})
 =B(\varphi,\psi,\mu,t)+(\tilde\varphi,P_{res}\psi_{\Lambda_+})e^{-i\mu t},\ t\geq 0,
\end{equation*}
and if we insert on the right hand side the explicit form of the projection $P_{res}$ we obtain
\begin{multline}
\label{semigroup_decomp_old_form}
 (\varphi,U(t)\psi)=B(\varphi,\psi,\mu,t)+\Vert\psi_\mu^{res}\Vert^{-2}(\tilde\varphi,\psi_\mu^{res})(\psi_\mu^{res},\psi_{\Lambda_+})e^{-i\mu t}=\\
 =B(\varphi,\psi,\mu,t)+\Vert\psi_\mu^{res}\Vert^{-2}(\tilde\varphi,\psi_\mu^{res})(\psi_\mu^{app},\psi)e^{-i\mu t},\quad t\geq 0.
\end{multline}
This is the form of the semigroup decomposition appearing in Ref. \cite{S2}. The term $B(\varphi,\psi,\mu,t)$ is a background term and the second term on the 
right hand side is the resonance term. The resonance state $\psi_\mu^{res}$, begin an eigenstate of the generator of the semigroup $\{Z_+(t)\}_{t\geq 0}$, 
determines the time evolution of the resonance term. Note that if the state $\psi$  is chosen to be orthogonal to $\psi_\mu^{app}$ then the resonance term in Eq. (\ref{semigroup_decomp_old_form}) vanishes. Hence the state $\psi_\mu^{app}$ is directly associated with the appearance of the resonance contribution on the right hand side of Eq. (\ref{semigroup_decomp_old_form}).
The reference to $\psi_\mu^{app}$ as an approximate resonance state follows from the fact that it can be shown that there is no choice of $\varphi$ and $\psi$ in 
Eq. (\ref{semigroup_decomp_old_form}) for which we obtain a pure resonance behavior, i.e., there is no choice of $\varphi$ and $\psi$ for which the background term disappears. In fact,
it can be shown that $\psi_\mu^{res}\in\mathcal H_{ac}\backslash(\mathcal H_{ac})_{\Lambda_+}$, i.e., $\psi_\mu^{res}$ is not in the range of $\Lambda_+$, the term
 $b(\psi;\mu)=(I-P_{res})\psi_{\Lambda_+}$ on the right hand side 
of Eq. (\ref{psi_lambda_plus_decomp_ver_2}) cannot disappear and the background term $B(\varphi,\psi,\mu,t)$ cannot be identically zero for any choice of $\varphi$ and $\psi$ (see Ref. \cite{S4}.
\par It is of particular interest to apply Eq. (\ref{semigroup_decomp_old_form}) to $\psi_\mu^{app}$. If we set in Eq. (\ref{semigroup_decomp_old_form}) $\varphi=\psi=\psi_\mu^{app}$ and use the fact that 
$\psi_\mu^{res}=\Lambda_+^{-1}\psi_\mu^{app}$ we get
\begin{equation*}
 \frac{(\psi_\mu^{app}, U(t)\psi_\mu^{app})}{\Vert\psi_\mu^{app}\Vert^2}=B_\mu(t)+e^{-i\mu t},
\end{equation*}
where $B_\mu(t):=\Vert\psi_\mu^{app}\Vert^{-2}B(\psi_\mu^{app},\psi_\mu^{app},\mu,t)$. It is shown in Ref. \cite{SHV} that in this case it is possible to obtain an estimate on the background term of the 
form
\begin{equation}
\label{background_estimate}
 \vert B_\mu(t)\vert\leq\left(\frac{\Vert\psi_\mu^{res}\Vert^4}{\Vert\psi_\mu^{app}\Vert^4}-1\right)^{1/2},\quad t\geq 0.
\end{equation}
(note that since $\psi_\mu^{app}=\Lambda_+\psi_\mu^{res}$ and since $\Lambda_+$ is contractive we always have $\Vert\psi_\mu^{app}\Vert\leq\Vert\psi_\mu^{res}\Vert$). For sharp (non-threshold) resonances, i.e., for 
resonance poles close to the real axis in the complex energy plane, the right hand side of the above inequality is small and hence the background term is small (see Ref. \cite{S4}).
\par Next, consider the approximate Lax-Phillips semigroup. By the intertwining relation in Eq. (\ref{z_+_intertwine}) we have
\begin{multline*}
 Z_{app}(t)=\Lambda_+U(t)\Lambda_- =Z_+(t)\Lambda_+\Lambda_- = Z_+(t)P_{res}+Z_+(t)(\Lambda_+\Lambda_- -P_{res})=\\
 =Z_{\text{\tiny QM}}(t)+Z_+(t)(\Lambda_+\Lambda_- -P_{res}).
\end{multline*}
Applying the approximate Lax-Phillips semigroup to the resonance state $\psi_\mu^{res}$ we obtain
\begin{multline*}
 Z_{app}(t)\psi_\mu^{res}=Z_{\text{\tiny QM}}(t)\psi_\mu^{res}+Z_+(t)(\Lambda_+\Lambda_- -P_{res})\psi_\mu^{res}=\\
 =e^{-i\mu t}\psi_\mu^{res} +Z_+(t)(\Lambda_+\Lambda_- -P_{res})\psi_\mu^{res}=e^{-i\mu t}\psi_\mu^{res} +Z_+(t)(\Lambda_+\Lambda_-\psi_\mu^{res}-\psi_\mu^{res}),
\end{multline*}
so that
\begin{equation}
\label{approx_LP_semigroup_eigenvector_deviation}
 \Vert Z_{app}(t)\tilde\psi_\mu^{res}-e^{-i\mu t}\tilde\psi_\mu^{res}\Vert\leq\Vert\tilde\psi_\mu^{res}-\Lambda_+\Lambda_-\tilde\psi_\mu^{res}\Vert,
\end{equation}
where $\tilde\psi_\mu^{res}=\Vert\psi_\mu^{res}\Vert^{-1}\psi_\mu^{res}$ is the normalized resonance state. We can now state our main result for this section
\begin{theorem}
\label{approx_proj_estimate}
Assume a scattering problem for which conditions (i)-(iv) hold and let $\psi_\mu^{res}$, $\tilde \psi_\mu^{res}$ and $\psi_\mu^{app}$ be as above. Let 
$\hat W_+^{\text{\tiny QM}}:\,\mathcal H_{ac}\mapsto L^2(\mathbb R_+)$ be the mapping to the outgoing energy representation for the problem and let $\psi_{\mu,+}^{app}\in L^2(\mathbb R_+)$ be given by 
$\psi_{\mu,+}^{app}(E)=[\hat W_+^{\text{\tiny QM}}\psi_\mu^{app}](E)$. Then we have the estimate
\begin{multline}
\label{approx_proj_ineq}
 \Vert\tilde\psi_\mu^{res}-\Lambda_+\Lambda_-\tilde\psi_\mu^{res}\Vert
 \leq C\left(1-\frac{\Vert\psi_\mu^{app}\Vert^2}{\Vert\psi_\mu^{res}\Vert^2}\right)^{1/2}\\
 +\left(\intop_0^\infty dE\,\left\vert1-\frac{E-\mu}{E-\overline\mu}\hat S_{\text{\tiny QM}}(E)\right\vert^2\,\frac{\vert\psi_{\mu,+}^{app}(E)\vert^2}
 {\Vert\psi_\mu^{app}\Vert^2}\right)^{1/2}.
\end{multline}
\hfill$\square$
\end{theorem}
\par Let us consider the two terms on the right hand side of Eq. (\ref{approx_proj_ineq}). In accord with the remark below Eq. (\ref{background_estimate}), if the resonance pole is at $z=\mu$ in the complex 
energy plane and if the resonance is narrow, i.e., if $\vert\text{Im}\,\mu\vert\ll 1$, then we have
\begin{equation*}
 1-\frac{\Vert\psi^{app}_\mu\Vert^2}{\Vert\psi^{res}_\mu\Vert^2}\ll 1.
\end{equation*}
Note that the first term on the right hand side of Eq. (\ref{approx_proj_ineq}) is small exactly when the estimate on the size of the background term in the evolution of the survival amplitude of 
$\psi_\mu^{app}$ in Eq. (\ref{background_estimate}) is also small. 
We see that the first term on the right hand side of Eq. (\ref{approx_proj_ineq}) measures the proximity of the resonance pole to the real axis, associated with the sharpness of the resonance at $z=\mu$. 
\par Turning to the second term on the right hand side of Eq. (\ref{approx_proj_ineq}) we recall that our assumption is that the S-matrix $\hat{\mathcal S}_{\text{\tiny QM}}(\cdot)$ has a simple pole at $z=\mu$. This leads us naturally to express $\hat S_{\text{\tiny QM}}(E)$ in the form $\hat S_{\text{\tiny QM}}(E)=\frac{E-\overline\mu}{E-\mu}\hat S_1(E)$. If the S-matrix 
$\hat{\mathcal S}_{\text{\tiny QM}}(E)$ were purely rational, i.e., $\hat S_{\text{\tiny QM}}(E)=\frac{E-\overline\mu}{E-\mu}$ (as is the case in the pure Lax-Phillips theory), we would have $1-\frac{E-\mu}{E-\overline\mu}\hat S_{\text{\tiny QM}}(E)=0$ and then the second term on the right hand side of Eq. (\ref{approx_proj_ineq}) would 
vanish. We conclude that the factor $\big\vert 1-\frac{E-\mu}{E-\overline\mu}\hat S_{\text{\tiny QM}}(E)\big\vert$ gives a measure of the deviation of the phase shift corresponding to the actual resonance at $z=\mu$ from the phase shift of an ideal, Breit-Wigner shaped, resonance associated with a purely rational S-matrix. In the second term on the right hand side of  
Eq. (\ref{approx_proj_ineq}) the factor $\big\vert 1-\frac{E-\mu}{E-\overline\mu}\hat S_{\text{\tiny QM}}(E)\big\vert$ is multiplied by the probability density function $\Vert\psi_\mu^{app}\Vert^{-2}\vert\psi_{\mu,+}^{app}(E)\vert^2$ which has a peak at the energy of the 
resonance and if the resonance is narrow then this peak is rather sharp. Hence the multiplication with the energy probability density of $\psi_\mu^{app}$ implies that the deviations of the factor 
$\big\vert 1-\frac{E-\mu}{E-\overline\mu}\hat S_{\text{\tiny QM}}(E)\big\vert$ from zero are evaluated in the vicinity of the resonance energy. We conclude that the second term on the right hand side of 
Eq. (\ref{approx_proj_ineq}) measures the deviation of the phase shifts of the resonance at $z=\mu$ from the phase shifts of an ideal resonance.
\par To summarize, to the resonance pole of the scattering matrix $\hat{\mathcal S}_{\text{\tiny QM}}(\cdot)$ at the point $z=\mu$ we assign a resonance state $\psi_\mu^{res}$. If the right hand side of the inequality in Eq. (\ref{approx_proj_ineq}) is small then $\psi_\mu^{res}$ is a good approximation to an eigenstate of the elements of the 
approximate Lax-Phillips semigroup $\{Z_{app}(t)\}_{t\geq 0}$ with eigenvalue $e^{-i\mu t}$. Thus, in the context of quantum mechanical resonance scattering and under the conditions that both terms on the right hand side of Eq. ({\ref{approx_proj_ineq}) are small, i.e., the resonance is sharp and exhibits phase shift which is close to that of an ideal resonance, 
Eq. (\ref{approx_LP_semigroup_eigenvector_deviation}) and Theorem \ref{approx_proj_estimate}, taken together, provide a result analogous to Theorem \ref{LP_main_result} of the Lax-Phillips theory.
\section{Proof of Theorem \ref{approx_proj_estimate}}
\label{proof_of_main_theorem}
\par Let $\hat W_+^{\text{\tiny QM}}:\,\mathcal H_{ac}\mapsto L^2(\mathbb R_+,\mathcal K)$ be the mapping to the outgoing energy representation corresponding to the scattering problem considered in 
Theorem \ref{approx_proj_estimate}. As mentioned in Section \ref{Lyapunov_op_trans_rep_in_LP_QM}, explicit expression for the mapping 
$\hat W_+^{\text{\tiny QM}}:\,\mathcal H_{ac}\mapsto L^2(\mathbb R_+,\mathcal K)$ is obtained by 
finding a complete set of outgoing solutions of the Lippmann-Schwinger equation. In this section we shall utilize the Dirac notation and denote an outgoing solution of the Lippmann-Schwinger equation 
corresponding to energy $E\in\mathbb R_+$ by $\vert E^-\rangle$ (recall that we assume that the a.c. spectrum is simple, i.e., the multiplicity of the a.c. spectrum is one, so that there are no degeneracy 
indices for the spectrum). The complete set of outgoing solutions is then $\{\vert E^-\rangle\}_{E\in\mathbb R_+}$. Similarly, 
If $\hat W_-^{\text{\tiny QM}}:\,\mathcal H_{ac}\mapsto L^2(\mathbb R_+,\mathcal K)$ is the mapping to the incoming energy representation then an explicit expression for this mapping 
is obtained by finding a complete set of incoming solutions of the Lippmann-Schwinger equation. Again we use the Dirac notation and denote an incoming solution of the Lippmann-Schwinger equation 
corresponding to energy $E\in\mathbb R_+$ by $\vert E^+\rangle$. The complete set of incoming solutions is then $\{\vert E^+\rangle\}_{E\in\mathbb R_+}$.
\par Recall that $\Lambda_+=M_+^{1/2}=(\hat W_+^{\text{\tiny QM}})^{-1}\Lambda_F (\hat W_+^{\text{\tiny QM}})$ and 
$\Lambda_- =M_-^{1/2}=(\hat W_-^{\text{\tiny QM}})^{-1}\Lambda_B (\hat W_-^{\text{\tiny QM}})$. We then have
\begin{multline*}
 \langle E^-\vert\Lambda_+\Lambda_-\vert E^{\prime -}\rangle=\intop_0^\infty dE_1\,\langle E^-\vert\Lambda_+\vert E_1^-\rangle\langle E_1^-\vert\Lambda_-\vert E^{\prime -}\rangle=\\
 =\intop_0^\infty dE_1\,\langle E^-\vert \left(\hat W_+^{\text{\tiny QM}}\right)^{-1}\Lambda_F \hat W_+^{\text{\tiny QM}}\vert E_1^-\rangle
 \langle E_1^-\vert \left(\hat W_-^{\text{\tiny QM}}\right)^{-1}\Lambda_B \hat W_-^{\text{\tiny QM}}\vert E^{\prime -}\rangle=\\
 =\intop_0^\infty dE_1\intop_0^\infty dE_2\intop_0^\infty dE_3\,\Lambda_F(E,E_1)\langle E_1^-\vert E_2^+\rangle\Lambda_B(E_2,E_3)\langle E_3^+\vert E^{\prime -}\rangle.
\end{multline*}
Now,
\begin{equation*}
 \langle E_1^-\vert E_2^+\rangle=\hat S_{\text{\tiny QM}}(E_1)\delta(E_1-E_2),\qquad \langle E_3^+\vert E^{\prime -}\rangle=\hat S_{\text{\tiny QM}}^*(E_3)\delta(E_3-E'),
\end{equation*}
where $\hat S_{\text{\tiny QM}}(\cdot)$ is the S-matrix. Therefore,
\begin{equation*}
 \langle E^-\vert\Lambda_+\Lambda_-\vert E^{\prime -}\rangle\intop_0^\infty dE_1\,\Lambda_F(E,E_1)\hat S_{\text{\tiny QM}}(E_1)\Lambda_B(E_1, E')\hat S_{\text{\tiny QM}}^*(E').
\end{equation*}
Applying this expression to $\psi_\mu^{res}$ we get
\begin{multline}
\label{psi_mu_ir_approx_proj}
 \langle E^-\vert\Lambda_+\Lambda_-\psi_\mu^{res}\rangle=\intop_0^\infty dE'\,\langle E^-\vert\Lambda_+\Lambda_-\vert E^{\prime -}\rangle
 \langle E^{\prime -}\vert\psi_\mu^{res}\rangle=\\
 =\intop_0^\infty dE'\intop_0^\infty dE_1\,\Lambda_F(E,E_1)\hat S_{\text{\tiny QM}}(E_1)\Lambda_B(E_1,E')\hat S_{\text{\tiny QM}}^*(E')\psi_{\mu,+}^{res}(E'),
\end{multline}
where $\psi_{\mu,+}^{res}(E'):=[\hat W_+^{\text{\tiny QM}}\psi_\mu^{res}](E)=\langle E^{\prime -}\vert\psi_\mu^{res}\rangle$. The transformation on the right hand side of Eq. (\ref{psi_mu_ir_approx_proj}) 
is between different functional representations of $\mathcal H_{ac}$, but it is understood that all of these transformations are acting in the function space $L^2(\mathbb R^+)$. In order to continue we 
write Eq. (\ref{psi_mu_ir_approx_proj}) in the form
\begin{multline}
\label{psi_mu_ir_approx_proj_2}
 \langle E^-\vert\Lambda_+\Lambda_-\psi_\mu^{res}\rangle
 =\intop_0^\infty dE'\intop_0^\infty dE_1\,\Lambda_F(E,E_1)\hat S_{\text{\tiny QM}}(E_1)(\Lambda_B-I)(E_1,E')\hat S_{\text{\tiny QM}}^*(E')
 \psi_{\mu,+}^{res}(E')\\
 +\intop_0^\infty dE'\,\Lambda_F(E,E')\psi_{\mu,+}^{res}(E')=\\
 =\intop_0^\infty dE_1\,\Lambda_F(E,E_1)\hat S_{\text{\tiny QM}}(E_1)[(\Lambda_B-I)\hat S_{\text{\tiny QM}}^*\psi_{\mu,+}^{res}](E_1)
 +\intop_0^\infty dE'\,\Lambda_F(E,E')\psi_{\mu,+}^{res}(E')=\\
 =[\Lambda_F\hat S_{\text{\tiny QM}}(\Lambda_B-I)\hat S_{\text{\tiny QM}}^*\psi_{\mu,+}^{res}](E)+[\Lambda_F\psi_{\mu,+}^{res}](E).
\end{multline}
where $I\equiv I_{L^2(\mathbb R_+)}$ is the identity operator on $L^2(\mathbb R_+)$. We will show that under appropriate conditions the norm 
$\Vert(\Lambda_B-I)\hat S_{\text{\tiny QM}}^*\psi_{\mu,+}^{res}\Vert_{L^2(\mathbb R^+)}$ is small.  Note first that by the positivity of $\Lambda_B$ we have the inequality
\begin{equation*}
 (\varphi,(I+\Lambda_B)\varphi)_{L^2(\mathbb R_+)}\geq \Vert\varphi\Vert^2_{L^2(\mathbb R_+)},\quad\forall\varphi\in L^2(\mathbb R^+),
\end{equation*}
by which we obtain that 
\begin{multline*}
 \Vert\varphi\Vert^2_{L^2(\mathbb R_+)}=(\varphi, (I+\Lambda_B)^{-1}(I+\Lambda_B)\varphi)_{L^2(\mathbb R_+)}=\\
 =((I+\Lambda_B)^{-1/2}\varphi,(I+\Lambda_B)(I+\Lambda_B)^{-1/2}\varphi)_{L^2(\mathbb R_+)}
 \geq((I+\Lambda_B)^{-1/2}\varphi,(I+\Lambda_B)^{-1/2}\varphi)_{L^2(\mathbb R_+)}=\\
 =(\varphi,(I+\Lambda_B)^{-1}\varphi)_{L^2(\mathbb R_+)},\quad\forall\varphi\in L^2(\mathbb R^+).
\end{multline*}
From this inequality we get that $\Vert(I+\Lambda_B)^{-1}\Vert_{L^2(\mathbb R_+)}\leq 1$. Noting that 
\begin{equation*}
 M_F+M_B=(P_{\mathbb R_+}\hat P_+P_{\mathbb R_+})\big\vert_{L^2(\mathbb R_+)}+(P_{\mathbb R_+}\hat P_-P_{\mathbb R_+})\big\vert_{L^2(\mathbb R_+)}
 =(P_{\mathbb R_+}(\hat P_+ +\hat P_-)P_{\mathbb R_+})\big\vert_{L^2(\mathbb R_+)}=I,
\end{equation*}
we have
\begin{equation*}
 M_F=I-M_B=(I+\Lambda_B)(I-\Lambda_B)\quad\Longrightarrow\quad I-\Lambda_B=(I+\Lambda_B)^{-1} M_F.
\end{equation*}
Thus,
\begin{multline*}
 \Vert(I-\Lambda_B)\hat S_{\text{\tiny QM}}^*\psi_{\mu,+}^{res}\Vert^2_{L^2(\mathbb R_+)}=\\
 =\Vert(I-\Lambda_B)^{1/2}(I-\Lambda_B)^{1/2}\hat S_{\text{\tiny QM}}^*\psi_{\mu,+}^{res}\Vert^2_{L^2(\mathbb R_+)}
 \leq\Vert(I-\Lambda_B)^{1/2}\hat S_{\text{\tiny QM}}^*\psi_{\mu,+}^{res}\Vert^2_{L^2(\mathbb R_+)}=\\
 =(\hat S_{\text{\tiny QM}}^*\psi_{\mu,+}^{res},(I-\Lambda_B)\hat S_{\text{\tiny QM}}^*\psi_{\mu,+}^{res})_{L^2(\mathbb R_+)}
 =(\hat S_{\text{\tiny QM}}^*\psi_{\mu,+}^{res},(I+\Lambda_B)^{-1}M_F \hat S_{\text{\tiny QM}}^*\psi_{\mu,+}^{res})_{L^2(\mathbb R_+)}=\\
 =(\hat S_{\text{\tiny QM}}^*\psi_{\mu,+}^{res},(I+\Lambda_B)^{-1}\Lambda_F^2 \hat S_{\text{\tiny QM}}^*\psi_{\mu,+}^{res})_{L^2(\mathbb R_+)}=\\
 =(\Lambda_F \hat S_{\text{\tiny QM}}^*\psi_{\mu,+}^{res},(I+\Lambda_B)^{-1}\Lambda_F \hat S_{\text{\tiny QM}}^*\psi_{\mu,+}^{res})_{L^2(\mathbb R_+)}
 \leq (\Lambda_F \hat S_{\text{\tiny QM}}^*\psi_{\mu,+}^{res},\Lambda_F \hat S_{\text{\tiny QM}}^*\psi_{\mu,+}^{res})_{L^2(\mathbb R_+)}=\\
 =(\hat S_{\text{\tiny QM}}^*\psi_{\mu,+}^{res},M_F \hat S_{\text{\tiny QM}}^*\psi_{\mu,+}^{res})_{L^2(\mathbb R_+)}.
\end{multline*}
In the derivation of this inequality we have used the fact that $[(I-\Lambda_B),\Lambda_F]=[I-(I-M_F)^{1/2},M_F^{1/2}]=0$. To summarize, we have
\begin{equation}
\label{I_minus_lambda_B_bound}
 \Vert(I-\Lambda_B)\hat S_{\text{\tiny QM}}^*\psi_{\mu,+}^{res}\Vert^2_{L^2(\mathbb R_+)}
 \leq (\hat S_{\text{\tiny QM}}^*\psi_{\mu,+}^{res},M_F \hat S_{\text{\tiny QM}}^*\psi_{\mu,+}^{res})_{L^2(\mathbb R_+)}.
\end{equation}
Plugging the definition $ M_F=P_{\mathbb R^+}P_+P_{\mathbb R^+}\big\vert_{L^2(\mathbb R^+)}$ into the inequality in Eq. (\ref{I_minus_lambda_B_bound}) we obtain
\begin{multline}
\label{I_minus_lambda_B_bound_2}
 \Vert(I-\Lambda_B)\hat S_{\text{\tiny QM}}^*\psi_{\mu,+}^{res}\Vert^2_{L^2(\mathbb R_+)}
 \leq (\hat S_{\text{\tiny QM}}^*\psi_{\mu,+}^{res},M_F \hat S_{\text{\tiny QM}}^*\psi_{\mu,+}^{res})=\\
 =(\hat S_{\text{\tiny QM}}^*\psi_{\mu,+}^{res},P_{\mathbb R^+}\hat P_+P_{\mathbb R^+} \hat S_{\text{\tiny QM}}^*\psi_{\mu,+}^{res})
 =(\hat S_{\text{\tiny QM}}^*\psi_{\mu,+}^{res},\hat P_+ \hat S_{\text{\tiny QM}}^*\psi_{\mu,+}^{res}).
\end{multline}
It is shown in Ref. \cite{SHV} that
\begin{equation*}
 \psi_\mu^{app}=\Lambda_+\psi_\mu^{res}=\intop_0^\infty dE\,\frac{1}{E-\mu}\vert E^-\rangle. 
\end{equation*}
so that in the outgoing energy representation we have $\psi_{\mu,+}^{app}(E)=\langle E^-\vert\psi_\mu^{app}\rangle=\frac{1}{E-\mu}$. Define
\begin{equation*}
 \psi_{\overline\mu}^{app}=\intop_0^\infty dE\,\frac{1}{E-\overline\mu}\vert E^+\rangle=\intop_0^\infty dE\,\vert E^+\rangle\,\psi_{\overline\mu,-}^{app}(E),
\end{equation*}
where $\psi_{\overline\mu,-}^{app}:=(E-\overline\mu)^{-1}$. Representing $\psi_{\overline\mu}^{app}$ in the outgoing energy representation we have
\begin{equation*}
 \psi_{\overline\mu,+}^{app}(E):=\langle E^-\vert\psi_{\overline\mu}^{app}\rangle
 =\intop_0^\infty dE'\,\frac{1}{E'-\overline\mu}\langle E^-\vert E^{\prime +}\rangle =\hat S_{\text{\tiny QM}}(E)\psi_{\overline\mu,-}^{app}(E).
\end{equation*}
Then, using Eq. (\ref{I_minus_lambda_B_bound_2}), we have
\begin{multline}
\label{I_minus_lambda_B_bound_3}
 \Vert(I-\Lambda_B)\hat S_{\text{\tiny QM}}^*\psi_{\mu,+}^{res}\Vert_{L^2(\mathbb R_+)}
 \leq\Vert \hat P_+\hat S_{\text{\tiny QM}}^*\psi_{\mu,+}^{res}\Vert_{L^2(\mathbb R_+)}\\
 \leq\Vert \hat P_+\hat S_{\text{\tiny QM}}^*(\psi_{\mu,+}^{res}-\psi_{\mu,+}^{app})\Vert_{L^2(\mathbb R_+)}
 +\Vert \hat P_+\hat S_{\text{\tiny QM}}^*(\psi_{\mu,+}^{app}-\psi_{\overline\mu,+}^{app})\Vert_{L^2(\mathbb R_+)}
 +\Vert \hat P_+\hat S_{\text{\tiny QM}}^*\psi_{\overline\mu,+}^{app}\Vert_{L^2(\mathbb R_+)}\\
 \leq\Vert\psi_{\mu,+}^{res}-\psi_{\mu,+}^{app}\Vert_{L^2(\mathbb R_+)}
 +\Vert\psi_{\mu,+}^{app}-\psi_{\overline\mu,+}^{app}\Vert_{L^2(\mathbb R_+)}
 +\Vert \hat P_+\hat S_{\text{\tiny QM}}^*\psi_{\overline\mu,+}^{app}\Vert_{L^2(\mathbb R_+)}\\
 \leq\Vert\psi_{\mu,+}^{res}-\psi_{\mu,+}^{app}\Vert_{L^2(\mathbb R_+)}+\Vert \psi_{\mu,+}^{app}-\hat S_{\text{\tiny QM}}\psi_{\overline\mu,-}^{app}\Vert_{L^2(\mathbb R_+)}
 +\Vert \hat P_+\psi_{\overline\mu,-}^{app}\Vert_{L^2(\mathbb R_+)}.
\end{multline}
We shall find an appropriate bound for each term on the right hand side of Eq. (\ref{I_minus_lambda_B_bound_3}). First, we have
\begin{multline*}
 \Vert\psi_{\mu,+}^{res}-\psi_{\mu,+}^{app}\Vert_{L^2(\mathbb R_+)}
 =\Vert\psi_\mu^{res}-\psi_\mu^{app}\Vert\leq\Vert(I+\Lambda_+)(\psi_\mu^{res}-\psi_\mu^{app})\Vert=\\
 =\Vert\psi_\mu^{res}+\Lambda_+\psi_\mu^{res}-\psi_\mu^{app}-\Lambda_+\psi_\mu^{app}\Vert
 =\Vert\psi_\mu^{res}+\psi_\mu^{app}-\psi_\mu^{app}-\Lambda_+\psi_\mu^{app}\Vert
 =\Vert\psi_\mu^{res}-\Lambda_+\psi_\mu^{app}\Vert.
\end{multline*}
Using Eq. (\ref{psi_lambda_plus_decomp_ver_2}) we get
\begin{multline*}
 \Lambda_+\psi_\mu^{app}=b(\psi_\mu^{app};\mu)+(\psi_\mu^{res},\Lambda_+\psi_\mu^{app})\Vert\psi_\mu^{res}\Vert^{-2}\psi_\mu^{res}=\\
 =b(\psi_\mu^{app};\mu)+(\Lambda_+\psi_\mu^{res},\psi_\mu^{app})\Vert\psi_\mu^{res}\Vert^{-2}\psi_\mu^{res}
 =b(\psi_\mu^{app};\mu)+\frac{\Vert\psi_\mu^{app}\Vert^2}{\Vert\psi_\mu^{res}\Vert^2}\psi_\mu^{res}
\end{multline*} 
so that
\begin{multline*}
 \Vert\psi_{\mu,+}^{res}-\psi_{\mu,+}^{app}\Vert_{L^2(\mathbb R_+)}
 \leq\Vert\psi_\mu^{res}-\Lambda_+\psi_\mu^{app}\Vert
 =\left\Vert\psi_\mu^{res}-\left(b(\psi_\mu^{app};\mu)+\frac{\Vert\psi_\mu^{app}\Vert^2}{\Vert\psi_\mu^{res}\Vert^2}\psi_\mu^{res}\right)\right\Vert\\
 \leq \left(1-\frac{\Vert\psi_\mu^{app}\Vert^2}{\Vert\psi_\mu^{res}\Vert^2}\right)\Vert\psi_\mu^{res}\Vert+\Vert b(\psi_\mu^{app};\mu)\Vert.
\end{multline*}
By Eq. (\ref{psi_lambda_plus_decomp_ver_2}) and by the orthogonality of $b(\psi_\mu^{app};\mu)$ and $\psi_\mu^{res}$ we obtain
\begin{equation*}
 \Vert\psi_\mu^{res}\Vert^2\geq\Vert\psi_\mu^{app}\Vert^2\geq\Vert\Lambda_+\psi_\mu^{app}\Vert^2
 =\Vert b(\psi_\mu^{app};\mu)\Vert^2+\frac{\Vert\psi_\mu^{app}\Vert^4}{\Vert\psi_\mu^{res}\Vert^4}\Vert\psi_\mu^{res}\Vert^2.
\end{equation*}
Hence
\begin{equation*}
 \Vert b(\psi_\mu^{app};\mu)\Vert^2\leq\left(1-\frac{\Vert\psi_\mu^{app}\Vert^4}{\Vert\psi_\mu^{res}\Vert^4}\right)\Vert\psi_\mu^{res}\Vert^2
 \leq 2 \left(1-\frac{\Vert\psi_\mu^{app}\Vert^2}{\Vert\psi_\mu^{res}\Vert^2}\right)\Vert\psi_\mu^{res}\Vert^2,
\end{equation*}
and finally
\begin{multline}
\label{psi_mu_ir_psi_mu_app_diff_est}
 \Vert\psi_{\mu,+}^{res}-\psi_{\mu,+}^{app}\Vert_{L^2(\mathbb R_+)}
 \leq \left(1-\frac{\Vert\psi_\mu^{app}\Vert^2}{\Vert\psi_\mu^{res}\Vert^2}\right)\Vert\psi_\mu^{res}\Vert
 +\sqrt{2}\left(1-\frac{\Vert\psi_\mu^{app}\Vert^2}{\Vert\psi_\mu^{res}\Vert^2}\right)^{1/2}\Vert\psi_\mu^{res}\Vert\\
 \leq (1+\sqrt{2})\left(1-\frac{\Vert\psi_\mu^{app}\Vert^2}{\Vert\psi_\mu^{res}\Vert^2}\right)^{1/2}\Vert\psi_\mu^{res}\Vert
\end{multline}
Next we obtain a convenient expression for the second term on the right hand side of Eq. (\ref{I_minus_lambda_B_bound_3}). For this we write the S-matrix 
$\hat S_{\text{\tiny QM}}(E)$ in the form
\begin{equation*}
 \hat S_{\text{\tiny QM}}(E)=\frac{E-\overline\mu}{E-\mu}\hat S_1(E),
\end{equation*}
and, according to the assumptions of Theorem \ref{approx_proj_estimate}, $\hat S_1(E)$ does not have a pole at $E=\mu$. We then have,
\begin{equation*}
 \hat S_{\text{\tiny QM}}(E)\psi_{\overline\mu,-}^{app}(E)=\frac{E-\overline\mu}{E-\mu}\hat S_1(E)\frac{1}{E-\overline\mu}
 =\hat S_1(E)\frac{1}{E-\mu}=\hat S_1(E)\psi_{\mu,+}^{app}(E),
\end{equation*}
Hence we get that
\begin{multline}
\label{psi_mu_app_minus_S_psi_mu_bar_est}
 \Vert\psi_{\mu,+}^{app}-\hat S_{\text{\tiny QM}}\psi_{\overline\mu,-}^{app}\Vert_{L^2(\mathbb R_+)}
 =\left(\intop_0^\infty dE\,\vert(1-\hat S_1(E)\vert^2\,\vert\psi_{\mu,+}^{app}(E)\vert^2\right)^{1/2}=\\
 =\left(\intop_0^\infty dE\,\left\vert1-\frac{E-\mu}{E-\overline\mu}\hat S_{\text{\tiny QM}}(E)\right\vert^2\,\vert\psi_{\mu,+}^{app}(E)\vert^2\right)^{1/2}.
\end{multline}
Finally, we obtain a convenient expression for the third term on the right hand side of Eq. (\ref{I_minus_lambda_B_bound_3}). We have
\begin{multline*}
 \Vert\hat P_+\psi_{\overline\mu,-}^{app}\Vert^2_{L^2(\mathbb R_+)}=(\psi_{\overline\mu,-}^{app},P_+\psi_{\overline\mu,-}^{app})_{L^2(\mathbb R_+)}=\\
 = -\frac{1}{2\pi i}\intop_0^\infty dE\,\intop_0^\infty dE'\,\frac{1}{E-\mu}\frac{1}{E-E'+i0^+}\,\frac{1}{E'-\overline\mu}=\\
 = \text{Re}\,\left[-\frac{1}{2\pi i}\intop_0^\infty dE\,\intop_0^\infty dE'\,\frac{1}{E-\mu}\frac{1}{E-E'+i0^+}\,\frac{1}{E'-\overline\mu}\right]=\\
 =\text{Re }\left[-\frac{1}{2\pi i}\intop_0^\infty dE\,\intop_0^\infty dE'\,\frac{E-\overline\mu}{\vert E-\mu\vert^2}
 \left(\frac{(E-E')-i\epsilon}{(E-E')^2+\epsilon^2}\right)\,\frac{E'-\mu}{\vert E'-\overline\mu\vert^2}\right]=\\
 = -\frac{1}{2\pi}\lim_{\epsilon\to 0^+}\text{Im }\left[\intop_0^\infty dE\,\intop_0^\infty dE'\,\frac{E-\overline\mu}{\vert E-\mu\vert^2}
 \left(\frac{(E-E')-i\epsilon}{(E-E')^2+\epsilon^2}\right)\,\frac{E'-\mu}{\vert E'-\overline\mu\vert^2}\right]=\\
 = -\frac{1}{2\pi}\lim_{\epsilon\to 0^+}\intop_0^\infty dE\,\intop_0^\infty dE'\,\frac{\text{Im }[(E-\overline\mu)((E-E')-i\epsilon)(E'-\mu)]}
 {\vert E-\mu\vert^2((E-E')^2+\epsilon^2)(\vert E'-\overline\mu\vert^2)}=\\
 = -\frac{1}{2\pi}\lim_{\epsilon\to 0^+}\intop_0^\infty dE\,\intop_0^\infty dE'\,\frac{(\text{Im}\,\overline\mu)\,(E-E')^2-\epsilon[(E-\text{Re}  
 \,\overline\mu)(E'-\text{Re}\,\overline\mu)+(\text{Im}\overline\mu)^2]}{\vert E-\mu\vert^2((E-E')^2+\epsilon^2)(\vert E'-\overline\mu\vert^2)}=\\
 = -\frac{\text{Im}\,\overline\mu}{2\pi}\left(\intop_0^\infty dE\,\frac{1}{\vert E-\mu\vert^2}\right)^2+\frac{1}{2}\intop_0^\infty dE\,\frac{1}{\vert E-\mu\vert^2}
 = \frac{1}{2}\left(1-\frac{\Vert\psi_{\mu,+}^{app}\Vert_{L^2(\mathbb R_+)}^2}{\Vert\psi_{\mu,+}^{res}\Vert^2_{L^2(\mathbb R_+)}}\right)=\\
 =\Vert\psi_{\mu,+}^{app}\Vert^2_{L^2(\mathbb R_+)}= \frac{1}{2}\left(1-\frac{\Vert\psi_\mu^{app}\Vert}{\Vert\psi_\mu^{res}\Vert}\right)\Vert\psi_\mu^{app}\Vert^2.
\end{multline*}
The last two equalities here follow from the fact that
\begin{equation*}
 \Vert\psi_\mu^{app}\Vert^2=\Vert\psi_{\mu,+}^{app}\Vert^2=\intop_0^\infty dE\,\frac{1}{\vert E-\mu\vert^2},\qquad 
 \Vert\psi_\mu^{res}\Vert^2=\Vert\psi_{\mu,+}^{res}\Vert^2=\intop_{-\infty}^\infty dE\,\frac{1}{\vert E-\mu\vert^2}=\frac{\pi}{\text{Im}\,\overline\mu}. 
\end{equation*}
(see Ref. \cite{S4}). Thus we get that
\begin{equation}
\label{P_plus_psi_mu_bar_est}
 \Vert \hat P_+\psi_{\overline\mu,-}^{app}\Vert_{L^2(\mathbb R_+)}
 =\frac{1}{\sqrt{2}}\left(1-\frac{\Vert\psi_\mu^{app}\Vert^2}{\Vert\psi_\mu^{res}\Vert^2}\right)^{1/2}\Vert\psi_\mu^{app}\Vert.
\end{equation}
Collecting the estimates in Eqns. (\ref{psi_mu_ir_psi_mu_app_diff_est}),(\ref{psi_mu_app_minus_S_psi_mu_bar_est}),(\ref{P_plus_psi_mu_bar_est}) 
and inserting in Eq. (\ref{I_minus_lambda_B_bound_3}) we find that
\begin{multline}
\label{I_minus_lambda_B_bound_4}
 \Vert(I-\Lambda_B)\hat S_{\text{\tiny QM}}^*\psi_{\mu,+}^{res}\Vert_{L^2(\mathbb R_+)}\\
 \leq\Vert\psi_{\mu,+}^{res}-\psi_{\mu,+}^{app}\Vert_{L^2(\mathbb R_+)}
 +\Vert \psi_{\mu,+}^{app}-\hat S_{\text{\tiny QM}}\psi_{\overline\mu,-}^{app}\Vert_{L^2(\mathbb R_+)}
 +\Vert \hat P_+\psi_{\overline\mu,-}^{app}\Vert_{L^2(\mathbb R_+)}\\
 \leq (1+\sqrt{2})\left(1-\frac{\Vert\psi_\mu^{app}\Vert^2}{\Vert\psi_\mu^{res}\Vert^2}\right)^{1/2}\Vert\psi_\mu^{res}\Vert
 +\left(\intop_0^\infty dE\,\left\vert1-\frac{E-\mu}{E-\overline\mu}\hat S_{\text{\tiny QM}}(E)\right\vert^2\,\vert\psi_{\mu,+}^{app}(E)\vert^2\right)^{1/2}\\
 +\frac{1}{\sqrt{2}}\left(1-\frac{\Vert\psi_\mu^{app}\Vert^2}{\Vert\psi_\mu^{res}\Vert^2}\right)^{1/2}\Vert\psi_\mu^{app}\Vert.
\end{multline}
In order to complete the proof we note that Eq. (\ref{psi_mu_ir_approx_proj_2}) lead to
\begin{multline*}
 \langle E^-\vert\Lambda_+\Lambda_-\psi_\mu^{res}\rangle-\langle E^-\vert\psi_\mu^{res}\rangle=\\
 =[\Lambda_F\hat S_{\text{\tiny QM}}(\Lambda_B-I)\hat S_{\text{\tiny QM}}^*\psi_{\mu,+}^{res}](E)
 +[\Lambda_F\psi_{\mu,+}^{res}](E)-\psi_{\mu,+}^{res}(E)=\\
  =[\Lambda_F\hat S_{\text{\tiny QM}}(\Lambda_B-I)\hat S_{\text{\tiny QM}}^*\psi_{\mu,+}^{res}](E)
 +\psi_{\mu,+}^{app}(E)-\psi_{\mu,+}^{res}(E),
\end{multline*}
and hence, using Eq. (\ref{I_minus_lambda_B_bound_4}) and Eq. (\ref{psi_mu_ir_psi_mu_app_diff_est}), we obtain
\begin{multline*}
 \Vert\psi_\mu^{res}-\Lambda_+\Lambda_-\psi_\mu^{res}\Vert
 \leq\Vert\Lambda_F\hat S_{\text{\tiny QM}}(\Lambda_B-I)\hat S_{\text{\tiny QM}}^*\psi_{\mu,+}^{res}\Vert_{L^2(\mathbb R_+)}
 +\Vert\psi_{\mu,+}^{app}-\psi_{\mu,+}^{res}\Vert_{L^2(\mathbb R^+)}\\
 \leq C \left(1-\frac{\Vert\psi_\mu^{app}\Vert^2}{\Vert\psi_\mu^{res}\Vert^2}\right)^{1/2}\Vert\psi_\mu^{res}\Vert
 +\left(\intop_0^\infty dE\,\left\vert1-\frac{E-\mu}{E-\overline\mu}\hat S_{\text{\tiny QM}}(E)\right\vert^2\,\vert\psi_{\mu,+}^{app}(E)\vert^2\right)^{1/2}\\
\end{multline*}
This last inequality, together with the fact that $\Vert\psi_\mu^{res}\Vert\geq\Vert\psi_\mu^{app}\Vert$, yields the desired result.
\par\hfill$\blacksquare$
\section{Conclusions}
\label{conclusions}
\par We have seen that in the context of quantum mechanical scattering it is possible to construct a structure analogous to the Lax-Phillips scattering theory. In fact, the main objects and representations 
of the Lax-Phillips theory are obtained as a particular example of a more universal construction. Let a scattering problem be defined on a Hilbert space $\mathcal H$ with a unitary evolution group
$\{U(t)\}_{t\in\mathbb R}$ having a self-adjoint generator $H$, such that $\sigma_{ac}(H)\not=\emptyset$, the multiplicity of the absolutely continuous spectrum is uniform and the mappings $\hat W_\pm\,:\,\mathcal H_{ac}\mapsto L^2(\mathbb R;\mathcal K)$ onto the incoming and outgoing spectral (energy) representations for $H$ exist. Denoting $P_{\sigma_{ac(H)}}\,:\,L^2(\mathbb R;\mathcal K)\mapsto L^2(\mathbb R;\mathcal K)$ the projection in $L^2(\mathbb R;\mathcal K)$ on the subspace $L^2(\sigma_{ac}(H);\mathcal K)$, we construct the operators $M_\pm(H)\,:\,\mathcal H_{ac}\mapsto\mathcal H_{ac}$ defined by 
\begin{equation*}
 M_\pm(H)=\hat W_\pm^{-1}P_{\sigma_{ac}(H)}\hat P_\pm P_{\sigma_{ac}(H)}\hat W_\pm
\end{equation*}
where $\hat P_\pm$ are, respectively, projections on the upper and lower half-plane Hardy spaces $\mathcal H^2_\pm(\mathbb R;\mathcal K)$. Then $M_\pm(H)$ are positive, contractive operators on $\mathcal H_{ac}$. We set
\begin{equation*}
 \Lambda_\pm(H):=M^{1/2}_\pm(H),
\end{equation*}
define
\begin{equation*}
 Z_{app}(t) :=\Lambda_+(H)U(t)\Lambda_-(H),
\end{equation*}
and obtain a generalized form of Eq. (\ref{lp_qm_analogy_ver_2}) in terms of the following list of correspondences between objects and representations of the Lax-Phillips case and the general case:
\begin{eqnarray}
\label{general_lp_qm_analogy}
 \underline{\text{LP scattering theory}} &{ }&  \underline{\text{QM scattering theory}}\notag\\
 U(t)=e^{-iKt} &\Longleftrightarrow& U(t)=e^{-iHt}\notag\\
 P_\pm &\Longleftrightarrow& \Lambda_\pm(H)\notag\\
 \psi(t)=P_+\psi(t)+P_+^\perp\psi(t) &\Longleftrightarrow& \psi(t)=\Lambda_+(H)\psi(t)+(I-\Lambda_+(H))\psi(t)\\
 \psi(t)=P_-^\perp\psi(t)+P_-\psi(t) &\Longleftrightarrow& \psi(t)=(I-\Lambda_-(H))\psi(t)+\Lambda_-(H)\psi(t)\notag\\
 Z_{\text{\tiny{LP}}}(t)=P_+U(t)P_-,\ t\geq 0 &\Longleftrightarrow& Z_{app}(t)=\Lambda_+(H)U(t)\Lambda_-(H),\ t\geq 0\notag\\
 \hat S_{\text{\tiny LP}}(E),\ \ E\in\mathbb R &\Longleftrightarrow& \hat S_{\text{\tiny QM}}(E),\ \  E\in\mathbb R^+\notag
\end{eqnarray}
The objects and representations on the left hand side of Eq. (\ref{general_lp_qm_analogy}), i.e., those of the Lax-Phillips theory, are obtained from those of the right hand side of 
Eq. (\ref{general_lp_qm_analogy}) in the particular case that $P_{\sigma_{ac}(H)}=I_{L^2(\mathbb R;\mathcal K)}$, i.e., in the case that $\sigma_{ac}(H)=\mathbb R$. For a scattering system satisfying 
assumptions (i)-(ii) in Section \ref{Lyapunov_op_trans_rep_in_LP_QM} we have $\sigma_{ac}(H)=\mathbb R_+$ and $P_{\sigma_{ac}(H)}=P_{\mathbb R_+}$ and the objects and representations 
on the right hand side of Eq. (\ref{general_lp_qm_analogy}) in this case are those listed on the right hand side of Eq. (\ref{lp_qm_analogy_ver_2}) (see the discussion preceding 
Eq. (\ref{lp_qm_analogy_ver_2}) in Section \ref{Lyapunov_op_trans_rep_in_LP_QM}). For a problem 
satisfying assumptions (i)-(ii) the results of Refs. \cite{S2,SSMH1,SSMH2} characterize $M_\pm$ as Lyapunov operators for the evolution, analogous to the Lyapunov operators $P_\pm$ of the Lax-Phillips theory, and 
the decompositions on the right hand sides of the third and fourth lines in Eq. (\ref{lp_qm_analogy_ver_2}) are, respectively, forward and backward transition representations for the evolution. If in addition to 
(i)-(ii) we assume that the scattering system satisfies (iii)-(iv) in Section \ref{res_poles_and_states_and_approx_LP_semigroup}, then the results of Refs. \cite{S3,S4,SHV} and 
Theorem \ref{approx_proj_estimate} of the present paper imply that to a resonance pole of the scattering 
matrix $\hat{\mathcal S}_{\text{\tiny QM}}(\cdot)$ at a point $z=\mu$, $\text{Im }\mu<0$, there is associated a resonance state $\psi_\mu^{res}\in\mathcal H_{ac}$ (or an eigensubspace in the more general case) 
which is an approximate eigenstate of the elements of the approximate Lax-Phillips semigroup $\{Z_{app}(t)\}_{t\geq 0}$. This last result is the analogue of Theorem \ref{LP_main_result}, a central result 
of the Lax-Phillips scattering theory associating with each pole of the Lax-Phillips scattering matrix $\mathcal S_{\text{\tiny LP}}(\cdot)$ a resonance state (or eigensubspace) in the Lax-Phillips Hilbert 
space $\mathcal H^{\text{\tiny LP}}$. The quality of the approximation to an exact semigroup behavior is quantified by the inequality in Eq. (\ref{approx_proj_ineq}). For an ideal resonance the two terms 
on the right hand side of Eq. (\ref{approx_proj_ineq}) vanish identically and the resonance state $\psi_\mu^{res}$ becomes an eigenstate of an exact semigroup.
\par The most obvious way in which the results of the present paper may be extended is by showing that the list of correspondences in 
Eq. (\ref {general_lp_qm_analogy}) is valid not only in terms of the formal construction of objects and representations but by proving that the objects and representations on the right hand side of 
Eq. (\ref{general_lp_qm_analogy}) have all the necessary properties beyond the case of a scattering system satisfying assumptions (i)-(iv) discussed in the present paper. This, together with an extension of 
Theorem \ref{LP_main_result} to an analogous result in the general case, would constitute a generalization of the Lax-Phillips theory into a formalism applicable to a much broader range of problems than 
those satisfying the strict assumptions of the original theory listed in Eq. (\ref{LP_assumptions}).
\end{document}